\documentclass[iop,apj]{emulateapj}
\usepackage{amsmath,amssymb}

\usepackage{xcolor}

\slugcomment{The Astrophysical Journal, 800:24 (20pp), 2015 February 10}

\usepackage[bookmarksnumbered,
colorlinks,
linkcolor=blue,
citecolor=black,
filecolor=black,
urlcolor=blue,
breaklinks=true,
]{hyperref}
\shortauthors{Knobel et al.}
\begin{document}
\title{Quenching of Star Formation in SDSS Groups:\\Centrals, Satellites, and Galactic Conformity}

\author{Christian~Knobel,
Simon~J.~Lilly,
Joanna~Woo,
and~Katarina~Kova\v{c}
}

\affil{Institute for Astronomy, ETH Zurich, CH-8093 Zurich, Switzerland}

\begin{abstract}
We re-examine the fraction of low redshift Sloan Digital Sky Survey satellites and centrals in which star formation has been quenched, using the environment quenching efficiency formalism that separates out the dependence of stellar mass. We show that the centrals of the groups containing the satellites are responding to the environment in the same way as their satellites (at least for stellar masses above $10^{10.3}\: M_\odot$), and that the well-known differences between satellites and the general set of centrals arise because the latter are overwhelmingly dominated by isolated galaxies. The widespread concept of ``satellite quenching'' as the cause of environmental effects in the galaxy population can therefore be generalized to ``group quenching''. We then explore the dependence of the quenching efficiency of satellites on overdensity, group-centric distance, halo mass, the stellar mass of the satellite, and the stellar mass and specific star formation rate (sSFR) of its central, trying to isolate the effect of these often interdependent variables. We emphasize the importance of the central sSFR in the quenching efficiency of the associated satellites, and develop the meaning of this ``galactic conformity'' effect in a probabilistic description of the quenching of galaxies. We show that conformity is strong, and that it varies strongly across parameter space. Several arguments then suggest that environmental quenching and mass quenching may be different manifestations of the same underlying process. The marked difference in the apparent mass dependencies of environment quenching and mass quenching which produces distinctive signatures in the mass functions of centrals and satellites will arise naturally, since, for satellites at least, the distributions of the environmental variables that we investigate in this work are essentially independent of the stellar mass of the satellite.\\[2mm]
\end{abstract}

\section{Introduction}

One of the key questions in the study of galaxies is how their evolution is influenced by their local environment, where environment can refer to any measurable parameter that somehow relates a galaxy to its cosmic neighborhood. Typically, environmental parameters characterize the neighborhood of a galaxy on scales similar or smaller than the size of its dark matter (DM) halo, since DM halos, being gravitational bound objects, can produce and maintain physical conditions within the halo which are very different from those outside the halo.\footnote{There are also indications that the state of a galaxy depends on environmental scales much larger than the virial radius \citep[e.g.,][]{lu2012}. For a general discussion of the concept of environment, we refer to, e.g., \cite{cooper2005}, Mo et al. (2010; Section 15.5)\nocite{mo2010}, \cite{carollo2013}, and \cite{haas2012}.} 

It is useful to differentiate between environmental parameters according to whether they are the same for all galaxies within a halo or whether they vary within the DM halo. Examples of the first category would be the halo mass, or the mass and the specific star formation rate (sSFR) of the central galaxy, while examples of the second category would be the local space density of galaxies or the distance of the galaxy from the group center.

One of the most striking events in the life of a galaxy is the relatively rapid cessation of its star formation rate (SFR), which leads to the observed bimodality within the population of galaxies \citep[e.g.,][]{strateva2001}\footnote{For an alternative view of quenching durations, see, e.g., \cite{schawinski2014} and \cite{woo2014}.}: most galaxies are either blue and star forming, with an SFR that is closely linked to the existing stellar mass \citep[e.g.,][]{elbaz2007, noeske2007, salim2007}, or red with an SFR that is lower by 1-2 orders of magnitude.  We refer to this cessation of star formation as ``quenching''.  Some of the basic questions are how the mechanisms of quenching are related to the stellar mass and the environment of the galaxies \citep[see, e.g.,][and references therein]{balogh2004, kauffmann2004, baldry2006, vandenbosch2008, peng2010, peng2012, presotto2012,wetzel2012,carollo2013,cibinel2013a,cibinel2013b,knobel2013,woo2013,carollo2014,bluck2014,omand2014,wetzel2014,woo2014}, and whether all galaxies are affected in the same way by these quantities, and particularly whether the central galaxies, which are typically the most massive galaxies lying at the center of a DM halo, respond to the environment in the same way as the surrounding satellite galaxies \citep[e.g.,][]{weinmann2006, vandenbosch2008, peng2012, knobel2013, kovac2014,tal2014}.

Analyzing the fraction of quenched galaxies for centrals and satellites as a function of stellar mass and environmental parameters has emerged as a fruitful way of gaining insights into the phenomenology and the physical processes of quenching.   However, one of the difficulties in interpreting this fraction is that many of the environmental parameters are statistically correlated both with each other or with the stellar mass of the galaxy. For example, there is good evidence that halo mass correlates with the mass of the central galaxy \citep[e.g.,][]{yang2009}, that the local overdensity correlates with the group-centric distance \citep[e.g.,][]{peng2012,woo2013}, and that the local overdensity also correlates with halo mass \citep[e.g.,][]{haas2012,carollo2013}. These correlations among the parameters can introduce spurious dependencies, i.e., dependencies which have no direct causal relation to quenching, if the quenched fraction is regarded as a function of certain parameters in isolation. Therefore, in principle one has to study the quenched fraction in the full parameter space and to vary only one parameter at a time, while keeping the others fixed (see, e.g., the discussion in \citealt{mo2010}; Section 15.5). However, this approach does also not necessarily guarantee that the measured dependence of the quenched fraction on a certain parameter is directly related to a physical quenching process, since the red fraction may also depends on the \textit{history} of the galaxy, i.e., how the galaxy is moving through the parameter space. Therefore, interpreting trends of the quenched fraction even in the full parameter space has turned out to be quite difficult and has led to some confusing and apparently inconsistent statements in the literature.  In this paper, we use the Sloan Digital Sky Survey (SDSS; \citealt{york2000}) and the group catalog of \cite{yang2012} to take a new look at some of the issues that have been recently raised. 

Several recent papers have constructed ``quenching efficiencies'' which construct the probability that a given galaxy is quenched relative to some comparison sample. An important example is the satellite quenching efficiency $\epsilon_{\rm sat}$ \citep{vandenbosch2008,peng2012,knobel2013,kovac2014,phillips2014,omand2014} which compares the quenched fraction of satellites at a given stellar mass to a comparison sample of central galaxies at the same stellar mass.  In simple terms, $\epsilon_{\rm sat}$ therefore reflects the chance that a given central galaxy becomes quenched when it falls into another DM halo, thereby becoming a satellite of another galaxy. The use of stellar mass alone in computing $\epsilon_{\rm sat}$ has been justified on the basis that the properties of central galaxies in general appear to be largely independent of environmental measures \citep{peng2012} (beyond those that correlate closely with stellar mass).  One of the goals of the present paper is to re-visit the environmental dependence of the properties of central galaxies, and particularly, of those central galaxies that are the centrals of the groups in which most of the satellites can be found rather than the general population of centrals.

Of course $\epsilon_{\rm sat}$ is simply a straightforward ``renormalization'' of the observable quenched fraction of satellite galaxies, and some analyses in the literature work instead with the latter quantity. The use of $\epsilon_{\rm sat}$, rather than the quenched fraction of satellites and centrals, is advantageous in our view since it removes the strong dependence of galaxy quenching on stellar mass \citep{peng2010, peng2012}. It also has a rather simple physical interpretation in terms of central galaxies becoming satellites (of the same stellar mass) when they enter another halo. This makes it easier to study the dependence of $\epsilon_{\rm sat}$ on other quantities, with stellar mass already taken into account. 

The view of field centrals as the set of representative progenitors of the satellites raises a few issues.  First, there is the question of the epoch at which the centrals and satellites should be compared.  \cite{wetzel2013} suggested that the satellites should be compared with centrals at the (significantly earlier) epoch at which the satellites first became satellites, and tried thereby to constrain the timescale for the quenching of satellites.    This approach is correct if one wishes to study the overall quenching of the satellites (as \citealt{wetzel2013} did), but would not be correct if satellites undergo separate continuing ``central-like'' quenching processes and if one wished to isolate the pure satellite effects. In this case one should compare the satellites with the field centrals at the epoch of observation and not at the epoch of infall.   
A further complication with using ``back-dated'' centrals is that in principle one should compare a given satellite with centrals at a slightly lower stellar mass, corresponding to the mass that the satellite had when it became a satellite.  This ``mass offset'' would depend on the time since infall (and thus possibly on other environmental parameters) but would be of order 0.15 dex for a star-forming SDSS satellite that became a satellite at $z = 0.3$.  Comparing centrals and satellites at the same epoch (as in this paper) would remove this ``mass offset'' for star-forming satellites, since star-forming centrals and satellites are observed to have broadly similar sSFR \citep{peng2012}.  However, and we thank the anonymous referee for pointing this out, the variation in star-formation histories that is observed between different satellites (i.e., between star-forming and quenched satellites) means that this ``mass offset'' will in any case vary from satellite to satellite and will be close to zero for any satellite that immediately quenched.  This makes it impossible to identify any single ``central-progenitor'' comparison mass for the satellites of a given observed stellar mass. This shuffling of progenitor masses due to different star-formation histories would also be accounted for in our approach, but only if the mix of star-formation histories was exactly the same for centrals and satellites, which of course we know is not the case, since satellites are more likely to be quenched at a given mass.  These are unlikely to be large effects, but together they caution against an over-interpretation of $\epsilon_{\rm sat}$.

Quite apart from these methodological issues, the real problem with the \cite{wetzel2013} approach is practical. There is continuing observational uncertainty about the rate of evolution of the quenched fraction of the centrals. A useful parameterization is the rate of change of the actively star-forming fraction with redshift, $df_{\rm SF}/dz = - df_{\rm q}/dz$.  \cite{wetzel2013} themselves used a strongly evolving fraction of quenched centrals with $df_{\rm SF}/dz \sim 0.6$ (and $\sim \! 0.65$ for all galaxies) which is much steeper than other estimates.    Most published estimates have not differentiated between centrals and satellites and consider the overall population.  However at high masses around $M^\ast$, the quenching of galaxies is dominated by the quenching of centrals and, since environmentally driven quenching becomes more important with time (see \citealt{peng2010}) the change in the quenched fraction of all galaxies should provide an upper bound to the change in the centrals alone. Estimates of this gradient over the redshift range of interest and for masses around $10^{10.5}\: M_\odot$ include 0.26 \citep[][their Figure 12]{moustakas2013}, 0.22 \citep[][their Figure 4]{george2013}, 0.17 \citep[][their Figure 8]{muzzin2013}. \nocite{tinker2013}Tinker et al.~(2013, their Figure 11) have 0.16 for COSMOS and 0.12 for PRIMUS, while four studies have zero or even negative values for this gradient \citep{ilbert2013,knobel2013,hartley2013,kovac2014}.  We note that the continuity analysis of \cite{peng2010} implies that if the Schechter $M^\ast$ is constant than the red fraction should also be more or less constant.   At the least, we conclude that it is hard to reliably correct for an evolving red fraction of centrals, and, given also the methodological issues discussed in the previous paragraph, we do not take this approach in the current paper.

To facilitate this sort of study, we introduce a new estimator for $\epsilon_{\rm sat}$, that can be applied to individual galaxies and thereby avoids the need for prior binning of samples in computing it.  Such binning introduces issues of matching between samples, unless the bins are extremely small, and becomes impracticable when the number of dimensions being examined becomes large.  The use of this new estimator enables us to flexibly examine the dependence of $\epsilon_{\rm sat}$ on a whole range of parameters of interest.  The environmental parameters which are considered in this paper are the halo mass, the mass of the central galaxy, the local galaxy density, the group-centric distance, and the sSFR of the central galaxy. 

The sSFR of the central galaxy emerges as a major factor in the quenching of satellites. This phenomenon was first noted --- at least in a large galaxy survey --- by \cite{weinmann2006} and was called by them ``galactic conformity''.  It has been studied in a number of subsequent papers \citep[e.g.,][]{ann2008,kauffmann2010,prescott2011,wang2012,kauffmann2013,hearin2013,phillips2014,hartley2014,hearin2014}.  The question of galactic conformity can be confusing, since similar effects can arise rather trivially. In this paper we attempt to clarify the meaning of this effect and the conditions under which it can arise. In addition we quantify its importance with respect to that of other environmental parameters.

Before concluding this introduction we briefly discuss the role of structure and morphology. Quenching has been observed to correlate strongly with morphology and galaxy structure in the local universe \citep{kauffmann2003b,franx2008,bell2008,vandokkum2011,robaina2012,cibinel2013b,mendel2013,schawinski2014,omand2014,bluck2014,woo2014}
and at high-$z$ \citep{wuyts2011,cheung2012,bell2012,szomoru2012,wuyts2012,barro2013,lang2014}.  On the other hand, quenched disk galaxies have also been observed in
significant numbers (\citealp{bundy2010,stockton2004,mcgrath2008,vandokkum2008,vandenbergh2009,vanderwel2011,salim2012,bruce2012,carollo2014}).

Despite this clear correlation between galactic structure and quenching, we do not consider galactic structure further in this paper, for the following reasons. First, it is clear that morphology can be trivially altered during quenching, e.g., through the fading of a star-forming disk enhancing the relative importance of the bulge \citep[see, e.g.,][]{carollo2014} and so some apparent correlation would be expected even if there was no physical link. Second, even a physical connection could have many different causal origins: quenching could result in structural changes, e.g., if quenching was caused by a starburst (or active galactic nucleus (AGN) activity) that had been triggered by a major merger \citep[see, e.g.,][]{dimatteo2005}, or structural changes could be responsible for quenching, e.g., if the presence of a massive bulge impedes star formation by stabilizing the disks \citep[see, e.g.,][]{martig2009,genzel2014} or by producing a massive black hole resulting in AGN driven winds \citep[see, e.g.,][]{fabian1999}. The structure of a galaxy could also be linked to its star formation history quite indirectly, e.g., if the density of a stellar population reflects its formation epoch \citep[see, e.g.,][]{carollo2013b}.  It is therefore not at all clear whether structure is the driver or the result of quenching. Finally, while the quenched fraction of satellites increases for denser environments, the early-type fraction of galaxies at constant mass \citep{bamford2009} and of quenched galaxies in particular \citep{carollo2014} is mostly constant with environment, indicating that mass and environment quenching either similarly affect galaxy morphology or are two reflections of the same physical phenomenon.

Our paper is organized as follows: in Section \ref{sec:data}, we describe the data and the derived data products that we use for our analysis. In Section \ref{sec:method}, we describe the methods of our analysis and, particularly, we introduce our new estimator for the satellite quenching efficiency. In Section \ref{sec:centrals_satellites}, we discuss, whether centrals and satellites are differently influenced by the environment, specifically differentiating between centrals in general (which are mostly singletons, i.e., centrals with no satellites within the sample) and those in the groups in which the sample of satellites reside. In Section \ref{sec:eps_sat_analysis}, we analyze the satellite quenching efficiency as a function of our environmental parameters --- with the exception of the sSFR of the central galaxy --- taking into account the correlations among these parameters. Section \ref{sec:galaxy_conformity} is then fully devoted to a discussion of galactic conformity i.e., the effect of the sSFR of the central. Finally, in Section \ref{sec:discussion} we discuss our results and Section \ref{sec:conclusion} concludes the paper.

Throughout this work, a concordance cosmology with $H_0 = 70\ \rm{km\; s^{-1}\; \rm{Mpc}^{-1}}$, $\Omega_{\rm m} = 0.3$, and $\Omega_\Lambda = 0.7$ is applied. All magnitudes are quoted in the AB system. We use the term ``dex'' to express the anti-logarithm, i.e., 0.1 dex corresponds to a factor $10^{0.1} \simeq 1.259$. We use $\log m$ as short term for $\log(m/M_\odot)$ and $\log$ sSFR as short term for $\log({\rm sSFR} \times {\rm Gyr})$. Throughout this paper all observational error bars indicate the 68\% confidence interval, which is estimated by means of 50 bootstrapped galaxy samples.

\section{Data and Data Products}\label{sec:data}

\subsection{Basic Sample}\label{sec:basic_section}

This work is based on the SDSS DR7 galaxy sample \citep{abazajian2009}. We use the Petrosian photometry and redshifts from the New York University Value-Added Galaxy Catalog \citep[NYU-VAGC;][]{blanton2005,padmanabhan2008}, where $K$-corrections for the estimation of absolute magnitudes and $V_{\rm max}$ volumes were obtained using the utilities (v4\_2) of \cite{blanton2007}. The stellar masses $m_{\ast}$ and SFR estimates used in this work correspond to an updated version of those derived in \cite{brinchmann2004}, where the uncertainty in the stellar mass is $\sim \! 0.1$ dex.\footnote{\url{http://www.mpa-garching.mpg.de/SDSS/DR7/}}

\begin{figure}
	\centering
	\includegraphics[width=0.48\textwidth]{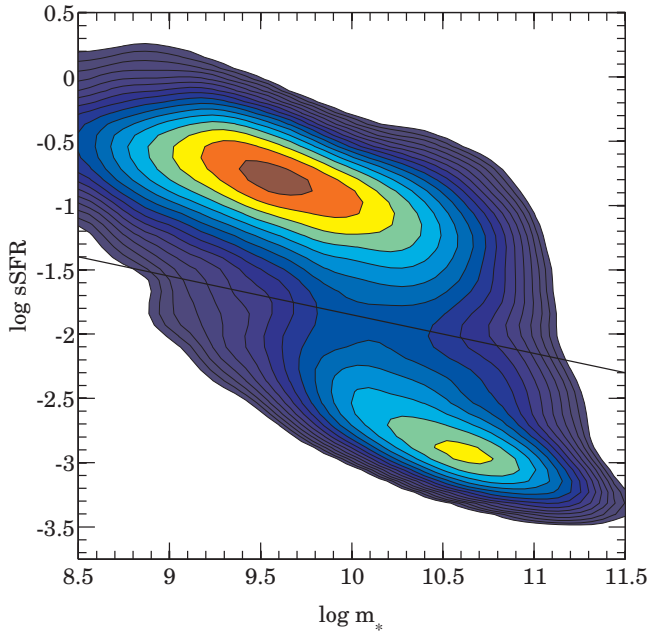}
	\caption{sSFR-$m_\ast$ diagram for the galaxies in our sample (including objects with $m_\ast < 10^9\:M_\odot$). The colors refer to the (unweighted) number density of galaxies in logarithmic scale. The black line, which is chosen to separate between star-forming sequence and quenched cloud, is given in Equation (\ref{eq:SSFR_division}).}\label{fig:SSFR_mass_diagram}
\end{figure}

We restrict our analysis to galaxies which lie in the redshift range $0.01 < z < 0.06$ and which have stellar masses $m_\ast > 10^9\:M_\odot$. Each galaxy is assigned a weight $W = 1/V_{\rm max} \times 1/$SSR, where SSR is the spectroscopic success rate of the galaxy (also obtained from the NYU-VAGC). The $V_{\rm max}$ was set to 1, if it exceeded the volume of our sample. The upper redshift limit was chosen so that our sample is essentially complete for all galaxies with $m_\ast > 10^{10}\:M_\odot$. After removing duplicated objects and objects with low quality redshifts, our sample contains $\sim\! 123,\! 000$ objects of which $\sim\! 53,\! 000$ objects have $m_\ast >10^{10}\:M_\odot$. In Figure \ref{fig:SSFR_mass_diagram} we show the sSFR-mass diagram, where the black line is given by
\begin{equation}\label{eq:SSFR_division}
\log {\rm sSFR} = -0.3 (\log m_\ast-10)-1.85\:.
\end{equation}
This dividing line was drawn by eye in order to separate the star-forming sequence from the quenched galaxies. 
Galaxies that lie above the line are regarded as star forming and those below as quenched. The numbers of star-forming and quenched galaxies for our sample are summarized in Table \ref{tab:galaxy_samples}.
\begin{deluxetable*}{cccccccc}
\tablewidth{0pt}
\tablecaption{The Galaxy and Group Sample}
\tablehead{
	\colhead{Richness} &
    \colhead{No.~of Groups} &
    	\colhead{Halo Mass}&
	 \multicolumn{2}{c}{No.~of Centrals\tablenotemark{a}} & 
	 \colhead{}  &
	 \multicolumn{2}{c}{No.~of Satellites\tablenotemark{a}} \\
	   \cline{4-5} \cline{7-8} \\
	   	\colhead{$\geq N_{\rm abs}$\tablenotemark{b}} &
	   		   	\colhead{} &
	   		   		   		\colhead{$\langle \log m_{\rm h} \rangle$\tablenotemark{c}} &
	 \colhead{Star Forming} & 
  \colhead{Quenched} &
  \colhead{} &
	\colhead{Star Forming} &
	\colhead{Quenched} 
  }
\startdata
$\phantom{00}0$     & 73801                             &  $11.52\pm 0.48 $ &    52306                          &    21495                       &&     (17471)    &   (14684)\\
$\phantom{00}1$     & 37464                             &  $11.90\pm 0.44$ &    18842                          &    18622                        &&     (16463)    &   (14615)\\
$\phantom{00}2$     &   $\phantom{0}$4526      &  $12.62\pm 0.55$ &      $\phantom{0}$1160   &    $\phantom{0}$3366   &&     (12418)    &   (13797)\\
$\phantom{00}3$     &   $\phantom{0}$1893      &  $13.02\pm 0.48$ &        $\phantom{00}$267   &    $\phantom{0}$1626   &&       $\phantom{0}$9645    &   12186\\
$\phantom{00}5$     &     $\phantom{00}$820    &  $13.39\pm 0.37$ &          $\phantom{000}$69   &    $\phantom{00}$751   &&       $\phantom{0}$7314    &   10380\\
$\phantom{0}10$     &     $\phantom{00}$336    &  $13.70\pm 0.28$ &     $\phantom{000}$19   &    $\phantom{00}$317   &&       $\phantom{0}$5214  & $\phantom{0}$8148\\
$\phantom{0}20$     &     $\phantom{00}$126    &  $13.97\pm 0.24$ &            $\phantom{0000}$4   &    $\phantom{00}$122   &&       $\phantom{0}$3177    &    $\phantom{0}$5762\\
$\phantom{0}50$     &       $\phantom{000}$30  &  $14.29\pm 0.14$ &      $\phantom{0000}$2   &    $\phantom{000}$28   &&       $\phantom{0}$1398 &$\phantom{0}$2996\\
$100$                      &         $\phantom{0000}$8 &  $14.43\pm 0.07$ &    $\phantom{0000}$1   &    $\phantom{0000}$7   &&         $\phantom{00}$556    &     $\phantom{0}$1306
\enddata
\tablenotetext{a}{The numbers refer to galaxies with $m_\ast > 10^{9}\: M_\odot$. Since we consider only satellites in groups with $N_{\rm abs}\geq 3$, we put the numbers of satellites for groups with $N_{\rm abs}< 3$ in parentheses.}
\tablenotetext{b}{Number of members with $m_\ast \geq 10^{10}\: M_\odot$.}
\tablenotetext{c}{The values correspond to the mean and the standard deviation of $\log m_{\rm h}$.}
\label{tab:galaxy_samples}
\end{deluxetable*}

\subsection{The Group Catalog}\label{sec:group_catalog}

The group catalog which we use in this work is taken from \cite{yang2012}.\footnote{\url{http://gax.shao.ac.cn/data/Group.html}} It was produced by the application of a sophisticated, iterative group-finding method, which is described in \cite{yang2007} in detail. We use the halo masses in that catalog that were derived by means of an abundance matching method between stellar mass and halo mass, which is based on the mass function of \cite{warren2006} and the transfer function of \cite{eisenstein1998}. Using simulated galaxy mock catalogs, \cite{yang2007} estimated that the uncertainties of the halo masses were $\lesssim \! 0.3$ dex, more or less independent of the luminosity of the groups.

A feature of the catalog of \cite{yang2007} is that essentially all galaxies are assigned to ``groups'' even if there is only one member. We refer to these as ``singletons''.  We define the richness $N$ of a given group to be the number of observed members within the parent galaxy sample. Since our sample is not mass complete below $10^{10}\:M_\odot$, we also introduce $N_{\rm abs}$ to be the number of members which are more massive than $10^{10}\: M_\odot$. To avoid biases in the selection of galaxies with redshift, we generally use $N_{\rm abs}$ to select groups (and singletons). However, it should be kept in mind that a group with $N_{\rm abs} =1$ or even $N_{\rm abs} =0$ may contain more than one galaxy below the mass completeness limit (see Table \ref{tab:galaxy_samples}).

The center of a group is defined by the $\log m_\ast$-weighted mean position of all its members. To compute the centers, we also consider galaxies with $m_\ast < 10^9\:M_\odot$, since the estimates become more robust the more galaxies we use and there is no danger here for selection biases with redshift. Some works define the center of a groups to be the position of the ``central'' galaxy, which is often taken to be the most massive within the group. This definition, however, is dependent on the identification of the central galaxy. If the central is identified wrongly, the estimated group center may be significantly off. The mass-weighted center has the advantage to be independent of the choice of the central galaxy and proves to be quite robust \citep[cf.~Figure 11 of][]{knobel2012}.

In order to compute the group-centric distance for each galaxy we first perform a principal component analysis of the projected positions of the group members on the sky to find the projected major and minor axes, $x$ and $y$, respectively, of the group. We then compute for each group member the coordinate difference $\Delta x_i = x_i - x_{\rm gr}$ and $\Delta y_i = y_i - y_{\rm gr}$ from the group center $(x_{\rm gr},y_{\rm gr})$ with respect to these axes, and subsequently we compute the rms $\Delta x_{\rm rms}$ and $\Delta y_{\rm rms}$ of $\Delta x_i$ and $\Delta y_i$, respectively, over all $N$ group members, which serves as an (asymmetric) measure of the extension of the group. Finally, the normalized ``group-centric distance'' $R$ for the galaxy $i$ is defined by
\begin{equation}\label{eq:R}
R_i = \sqrt{\left(  \frac{ \Delta x_i}{\Delta x_{\rm rms}}\right)^2  + \left( \frac{ \Delta y_i}{\Delta y_{\rm rms}}\right)^2 }\:.
\end{equation}
The advantage of this definition is that it can be directly computed from the positional information of the galaxies relative to each other and that it takes into account that, in general, galaxy groups are not rotationally symmetric.  It works well for groups with at least five identified members, which comprise 43\% of the groups and 81\% of the satellites used in this paper (as defined below).  For the remaining groups  (i.e., those with $N_{\rm abs} \geq 3$ and $N < 5$) we simply set $x$ and $y$ to be the right ascension $\alpha$ and declination $\delta$, respectively. We, however, checked that all of our results do not change, if we use instead the group-centric distance normalized by the virial radius (as, e.g., in \citealt{woo2013}).

The galaxy population of the groups is often divided into ``centrals'' and ``satellites'', where the centrals are thought to be the most massive galaxies within the groups lying at the deepest point of the gravitational potential well of the halo and the satellites are just all remaining group members. This framework is sometimes called the ``central galaxy paradigm'' \citep{vandenbosch2005}. However, there are indications that reality may be more complex than this \cite[cf.][]{skibba2011} and there are also circumstances when there may be groups for which no well-defined central should exist (e.g., in the case of merging or unrelaxed groups, see extensive discussion in \citealt{carollo2013}). In practice, the identification of the central galaxy is further exacerbated due to observational uncertainties in the stellar mass as well as imperfections in the group membership due to misidentification of groups (``fragmentation'' or ``over-merging'' of groups, the presence of unrelated interlopers or the exclusion of actual members; cf., e.g., \citealt{knobel2009}).

For our analysis we simply define the centrals to be the most massive galaxies within the groups and the satellites to be all remaining group members.  However, to avoid likely cases of misclassification and to enhance the fidelity of both centrals and satellites, we also require the central to have a group-centric distance $R < 2$, otherwise the whole group is discarded.  This selection is aimed to exclude groups which are either completely unrelaxed or which have a massive interloper at the outskirts \citep[][]{carollo2013,cibinel2013a}. In addition, since the purity of satellites is naturally worse for small groups (i.e., particularly for pairs) and at the outskirts of groups \citep[cf., e.g.,][]{knobel2009,knobel2012}, we also require that our sample of satellites lie in groups with $N_{\rm abs}\geq 3$ and consider only group-centric distances $R < 3$.  

We do not attempt to quantify, or correct for, the resulting purity and completeness of the central-satellite classification \citep[cf., e.g.,][]{knobel2013}, because the impurities of both centrals and satellites that are expected within the catalog from \cite{yang2012} are $\lesssim 10\%$ for most stellar masses and galaxy densities \citep{hirschmann2014}. The correct identification of the central is particularly important for studies of galactic conformity, which is based on whether the central is quenched or not.  Misidentification of the central can cause all the associated satellites to be moved to the incorrect bin. Therefore, for our conformity analyses we use an additional ``high purity sample'' as a check to estimate the impact of such misidentifications. Since the uncertainty of the stellar mass is $\lesssim\! 0.1$ dex, we require for this high purity sample that all galaxies with a stellar mass difference of less than $0.1$ dex from the nominal central must be all star forming or all  quenched. By discarding all groups which do not meet this condition, we remove 5\% of all satellites that are in groups with $N_{\rm abs}\geq 3$ and 15\% of all satellites that are in groups with $N_{\rm abs}\geq 20$. In order to not enforce a spurious conformity signal on our high purity sample, we also discard the second, third, etc. ranked objects in stellar mass, which are within this 0.1 dex difference from the first ranked object and which were required to show the same star formation behavior as the central.

Within our sample there are two particular clusters that are rich and have a star forming central. These will be treated specially. The first of these clusters has $N_{\rm eff} = 104$ and a central with $\log m_\ast \simeq 11.8$ and log sSFR $\simeq -2.24$, while the second one has $N_{\rm eff} = 65$ and a central with $\log m_\ast \simeq 11.1$ and log sSFR $\simeq -0.68$. In the first case, the classification of the central as star forming is uncertain because it lies only 0.14 dex above the black line in Figure \ref{fig:SSFR_mass_diagram} and therefore falls between the star forming sequence and the quenched cloud. In the second case, the central lies in the star forming sequence and is even included in our high purity sample, which was introduced in the previous paragraph. Since these clusters contain many satellites, they can dominate the population of satellites with a star forming central in certain bins of our environmental parameters. Therefore, since these clusters constitute rare objects which are not representative for the bulk of groups in our sample, we exclude them from our sample, whenever they affect the result substantially.

\subsection{The Density Field}\label{sec:density_field}

\begin{figure}
	\centering
	\includegraphics[width=0.48\textwidth]{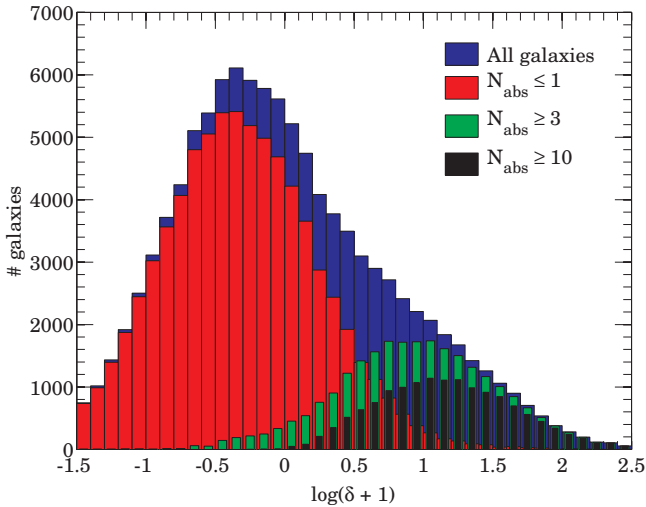}
	\caption{Distributions of galaxy overdensity $\delta$ (see Section \ref{sec:density_field}). The color refers to different richnesses $N_{\rm abs}$ as indicated in the legend. The red histogram consists mostly of centrals, while the green and the black histograms consist mostly of satellites.}\label{fig:density_field_histogram}
\end{figure}

Following \cite{peng2010,peng2012} and \cite{woo2013} we estimate the local galaxy overdensity $\delta = (n - \bar n)/\bar n$, where $n$ is the number density of galaxies which we use as tracers and $\bar n$ the corresponding mean number density, by means of the ``fifth nearest neighbor'' approach (for a general discussion on density fields see, e.g., \citealt{kovac2010}). That is, around each galaxy $i$, we put a cylinder with radius $r$ perpendicular to the line of sight and length $\pm dz$ along the line of sight, where $dz =  (1+z_i) dv/c$ with $dv = 1000$ km s$^{-1}$ and $c$ the speed of light. We then continuously shrink $r$ until the cylinder contains exactly five tracer galaxies (if the galaxy $i$ is a tracer galaxy, then it is counted as well). To guarantee that the density field is not subjected to biases with redshift, we use only a mass complete sample of tracers (i.e., galaxies with $m_\ast > 10^{10}\: M_\odot$). If $V_i$ is the volume of the cylinder around galaxy $i$, we regard $n_i = 1/V_i$ as the estimate of the local number density of galaxies. However, since these densities $n_i$ are centered around galaxies, i.e., special locations, they cannot be directly used to estimate the mean density $\bar n$. Instead, to estimate $\bar n$, we put the centers of the cylinders at randomly sampled positions, which have the same distribution with respect to the right ascension $\alpha$, the declination $\delta$, and the redshift $z$ as the actual galaxies. We construct realizations of such random positions by scrambling the redshifts among the galaxies, while keeping their $\alpha$ and $\delta$ fixed. Finally, we can compute the fifth nearest neighbor overdensity at the position of the $i$th galaxy by $\delta_i = (n_i-\bar n)/\bar n$. The distributions of $\log(\delta + 1)$ for group galaxies with various $N_{\rm abs}$ are shown in Figure \ref{fig:density_field_histogram}. As expected the higher $N_{\rm abs}$ the higher is the average $\log(\delta + 1)$ of the corresponding group galaxies.

As was pointed out in the literature \citep[e.g.,][]{peng2012,carollo2013,woo2013,kovac2014} a major disadvantage of the fifth nearest neighbor approach is that for small groups (i.e., groups with $N_{\rm abs} \lesssim 5$) the length scale, on which $\delta$ is estimated, can be much larger than the size of the corresponding DM halo. That is, for galaxies within such groups $\delta$ is not probing the local density of tracer galaxies within the halo, but rather the density on a scale beyond the group, which furthermore depends strongly on the richness of the group. For such low richness groups (and singletons), $\delta$ defines an environment that is expected to have little influence on the quenching of galaxies. For much richer groups, $\delta$ in contrast is measuring the environment well within the group virial radius. Therefore, we should not be surprised, if we do not find a dependence of the quenched fraction on $\delta$ for low richness groups. 

\section{Methods}\label{sec:method}

In this section we describe the methods that we use for our analysis. In the following, an index $i$ refers to the $i$th galaxy in the sample. For example, $m_{\ast,i}$ is the stellar mass of the $i$th galaxy and for any generic parameters $p_1,p_2,\ldots$ the symbols $p_{1,i}, p_{2,i},\ldots$ denotes the position of the $i$th galaxy in the corresponding multi-dimensional parameter space.

\subsection{Estimation of $f_{\rm q}$}

The fraction $f_{\rm q | S}$ of quenched objects within a sample $S$ is estimated by
\begin{equation}\label{eq:f_q}
f_{\rm q | S} = \frac{\sum_i W_i q_i}{\sum_i W_i}\:,
\end{equation}
where $q_i$ is 1 if the galaxy $i$ is quenched and 0 otherwise. The sum runs over all galaxies within the sample $S$ and $W_i$ are the weights of the galaxies in the sample as described in Section \ref{sec:basic_section}.  The sample $S$ usually refers to a (sub)sample of centrals (i.e., $S = \rm cen$) or satellites (i.e., $S = \rm sat$). By selecting galaxies within bins of some given parameters $p_1, p_2,\ldots$ (e.g., stellar mass $m_\ast$, overdensity $\delta$, halo mass $m_{\rm h}$, etc.), it is straightforward to compute  $f_{\rm q | S}(p_1,p_2,\ldots)$ as a function of these parameters.

\subsection{Estimation of $\epsilon_{\rm sat}$}\label{sec:eps_sat}

The estimation of the satellite quenching efficiency $\epsilon_{\rm sat}$ is a bit more subtle. In the literature \citep{vandenbosch2008,peng2012,knobel2013,kovac2014},  $\epsilon_{\rm sat}$ was computed either as a function of $m_\ast$ only or as a function of $m_\ast$ and $\delta$ as follows:
\begin{equation}\label{eq:eps_sat_standard}
\epsilon_{\rm sat}(m_\ast,\delta) = \frac{f_{\rm q | sat}(m_\ast,\delta) - f_{\rm q | cen}(m_\ast)}{1 - f_{\rm q | cen}(m_\ast)}\:.
\end{equation}
It should be noted that in either case the quenched fraction of centrals, which plays the role of the background or control sample, is only considered as a function of $m_\ast$ and not as a function of $\delta$.  Defined like this, $\epsilon_{\rm sat}(m_\ast,\delta)$ loosely has the simple interpretation as the probability that a central becomes quenched, as it falls into another DM halo becoming a satellite \citep[for a discussion see][]{knobel2013,kovac2014}. The special role of the stellar mass $m_\ast$ is that it is the one quantity that should stay the same (or more strictly vary continuously) when a central becomes a satellite, since $m_\ast$ is an intrinsic rather than an environmental parameter.   As noted above, the quenching probability depends strongly on mass in a similar way for centrals and satellites, as also evidenced by their similar Schechter $M^*$ characteristic mass \citep[see][]{peng2012}. By using the mass-dependent quenched fraction of centrals to ``renormalize'' the quenched fraction of the satellites, we effectively take out this strong empirical mass-dependence.

Because centrals and satellites are distributed differently in the $(m_\ast,\delta)$ parameter space, estimating $\epsilon_{\rm sat}(m_\ast,\delta)$ requires the use of very small bins or a careful matching of the distributions of both samples to each other \citep[][]{kovac2014}. In order to circumvent this computationally inconvenient situation and to allow us to examine the dependence of $\epsilon_{\rm sat}$ on any set of parameters $p_1, p_2, \ldots$, we introduce a new estimator that defines a quenching efficiency on a galaxy-by-galaxy basis such that the quenching efficiency for a set of galaxies is obtained by simply averaging over these individual quenching efficiencies.

We must first compute the quenched fraction of centrals as a function of their stellar mass, using Equation (\ref{eq:f_q}), i.e.,
\begin{equation}
f_{\rm q | cen} = \frac{\sum_i W_{i} q_{i}}{\sum_i W_{i}}\:,
\end{equation}
where, as above, $q_i$ is unity for quenched galaxies and zero otherwise.  Since there are usually many more centrals than satellites, $f_{\rm q | cen}$ can be evaluated on a fine grid in stellar mass, and represented by some suitably smooth curve, which has to be computed only once. Then, by inserting
\begin{equation}
f_{\rm q | sat} = \frac{\sum_i W_{i} q_{i}}{\sum_i W_{i}}
\end{equation}
into (\ref{eq:eps_sat_standard}), where the sum runs over all satellites within a given bin $(m_\ast,\delta)$, we obtain
\begin{align}
\epsilon_{\rm sat}(m_\ast,\delta) &= \frac{ \dfrac{\sum_i W_{i} q_{i}}{\sum_i W_{i}}-f_{\rm q | cen}(m_\ast)}{1-f_{\rm q | cen}(m_\ast)}\\
 &= \frac{\sum_i W_{i}\left( \dfrac{ q_{i} - f_{\rm q | cen}(m_\ast)}{1-f_{\rm q | cen}(m_\ast)} \right) }{\sum_i W_{i}}\:,
\end{align}
where for the second equality we have just rearranged the terms.  If we now shrink the size of the bin until only one satellite is left, we can effectively interpret the corresponding value, i.e.,
\begin{equation}\label{eq:eps_sat_i} 
\epsilon_{{\rm sat},i}
= \frac{q_i - f_{{\rm q | cen}}(m_{\ast,i})}{1-f_{{\rm q | cen}}(m_{\ast,i})}\:,
\end{equation}
as an estimate of the satellite quenching efficiency for the single galaxy $i$. Here, $f_{{\rm q | cen}}(m_{\ast,i})$ is our previously computed smooth curve for the quenched fraction of centrals, which is evaluated at $m_{\ast,i}$. Of course, the meaning of $\epsilon_{{\rm sat},i}$ for any individual galaxy is limited, because it is either quenched or not, i.e., $q_i$ is either unity or zero.  However, we can obtain $\epsilon_{\rm sat}$ for any set of galaxies by simply averaging over these individual estimates, i.e.,
\begin{equation}\label{eq:epsilon_sat_new}
\epsilon_{\rm sat}(m_\ast,\delta) = \frac{\sum_i W_{i}\epsilon_{{\rm sat},i}}{\sum_i W_{i}}\:,
\end{equation}
where the sum runs over all satellites within a particular sample of interest. 

Estimating $\epsilon_{\rm sat}$ using Equation (\ref{eq:epsilon_sat_new}) not only avoids the need for sample matching, but also enables us to easily compute $\epsilon_{\rm sat}(p_1,p_2,\ldots)$ as a function of as many parameters $p_1,p_2,\ldots$ as we like.  The individual estimates $\epsilon_{{\rm sat},i}$ need only be calculated once for each $i$th galaxy and then combined at will to compute $\epsilon_{\rm sat}$ for any set of galaxies and thus also to compute $\epsilon_{\rm sat}(p_1,p_2,\ldots)$.

For some applications it could be useful to generalize the definition of the satellite quenching efficiency in Equation (\ref{eq:eps_sat_standard}) to include additional parameter dependencies in the quenched fraction of centrals, i.e., to use $f_{\rm q | cen}(m_\ast,\tilde p_1,\tilde p_2,\ldots)$ instead of $f_{\rm q | cen}(m_\ast)$. This generalization changes the interpretation of the satellite quenching efficiency in so far as it now corresponds to the probability that a central with mass $m_\ast$, which originally resided in the environment $\tilde p_1,\tilde p_2,\ldots$, becomes quenched, as it becomes a satellite in the environment $p_1, p_2,\ldots$. We call the parameters $\tilde p_1,\tilde p_2,\ldots$ ``secondary parameters'' and denote them by a tilde to distinguish them from the primary parameters $p_1,p_2,\ldots$. Using our estimator of the previous paragraph, this generalized satellite quenching efficiency can be easily computed by simply exchanging $f_{\rm q | cen}(m_{\ast,i})$ in Equation (\ref{eq:eps_sat_i}) by $f_{\rm q | cen}(m_{\ast,i},\tilde p_{1,i},\tilde p_{2,i},\ldots)$.  If we consider the dependencies $f_{\rm q | cen}(m_\ast,\tilde p_1,\tilde p_2,\ldots)$ we denote the corresponding satellite quenching efficiency by $\epsilon_{{\rm sat},\tilde p_1,\tilde p_2,\ldots}(p_1,p_2,\ldots)$. The mass $m_\ast$ is not explicitly included in the notation, since we assume here that it is always considered in $f_{\rm q | cen}$. We will use this special satellite quenching efficiency only once below, in Section \ref{sec:eps_sat_analysis}.  The physical meaning of this more general $\epsilon_{{\rm sat},\tilde p_1,\tilde p_2,\ldots}(p_1,p_2,\ldots)$ is to more directly compare the properties of a given satellite with those that a central has in exactly the same location in the $(\tilde p_1,\tilde p_2,\ldots)$ space.

\subsection{Matching of Galaxy Samples}

To exclude spurious segregation effects when comparing the $\epsilon_{\rm sat}$ of two different samples of satellites, using the above individual estimator $\epsilon_{{\rm sat},i}$, we must still worry whether these two samples are well matched with respect to one or more parameters, i.e., whether the distribution of galaxies within the bin in question is the same in the two samples  \citep[][]{kovac2014}. The same is also true when we compute a quenched fraction for two samples.  We perform the required matching as follows: if the samples are to be matched in the parameters $p_1,p_2,\ldots$, we select for each galaxy $i$ of the smaller sample the corresponding galaxy $j$ of the larger sample which has the smallest separation to the galaxy $i$ in the considered parameter space. If this separation is smaller than $0.1\sqrt{N_{\rm p}}$ in a log space, where $N_{\rm p}$ is the number of matching parameters, i.e., if it holds
\begin{equation}
\sqrt{(p_{1,i}-p_{1,j})^2 + (p_{2,i}-p_{2,j})^2 +\cdots } < 0.1 \sqrt{N_{\rm p}}\:,
\end{equation}
the galaxy pair $(i,j)$ constitutes a match and the galaxy $j$ is removed from the sample, so that each galaxy of the larger sample is considered only once. If the separation is larger than $0.1\sqrt{N_{\rm p}}$, then there is no match and the galaxy is removed from the smaller sample. The matching for parameters is always performed in logarithmic space, since we will find that the relevant parameter ranges in logarithmic space are fairly similar for different parameters (see Section \ref{sec:eps_sat_analysis}). After applying this procedure for each galaxy $i$ of the smaller sample we end up with two new, ``matched'' subsamples which contain the same number of galaxies, are paired galaxy by galaxy and which should therefore have a very similar distribution across the bin.

\section{Environmental quenching of centrals}\label{sec:centrals_satellites}

There is an ongoing discussion as to what extent the centrals of galaxy groups are `special' compared with satellites \citep[for recent discussions based on ``brightest group galaxies'' see, e.g.,][]{vonderlinden2007,shen2014}. In the context of galaxy quenching, \cite{peng2012} made the hypothesis that, while satellites encounter both mass quenching and environment quenching, centrals encounter only mass quenching, and are not influenced by their local environment. There were two pieces of evidence which led \cite{peng2012} to this conclusion: first, it was shown that the red fraction of the bulk of centrals at fixed mass is much more weakly dependent on $\delta$ than the corresponding red fraction of satellites. The weak deviations that were observed could be attributed to secondary effects (e.g., impurities of the group sample, especially for low mass centrals in high density regions). Second, it was pointed out that the mass function of red centrals, unlike that of red satellites, does not show any signs of a second Schechter function at the low mass end.  In  \cite{peng2010} it was clear that this second Schechter component, with a faint end slope that is essentially the same as that of the star-forming population, arises from the action of the (mass-independent) ``environment quenching''.

Both arguments are still true. It should however be appreciated that, as is clear from the figures and discussion in \cite{peng2012}, the range of $\delta$ of the majority of centrals is different from that of the majority of satellites --- simply because most of centrals are not the centrals of the groups that contain most satellites and are in fact singletons.   In this section, we will investigate how the quenching of centrals is influenced by the environment by looking at the same regions in the multi-parameter space that are occupied by both centrals and satellites, i.e., by isolating those centrals that are the centrals of the richer groups, $N_{\rm abs} \geq 3$.

\begin{figure*}
	\centering
	\includegraphics[width=0.7\textwidth]{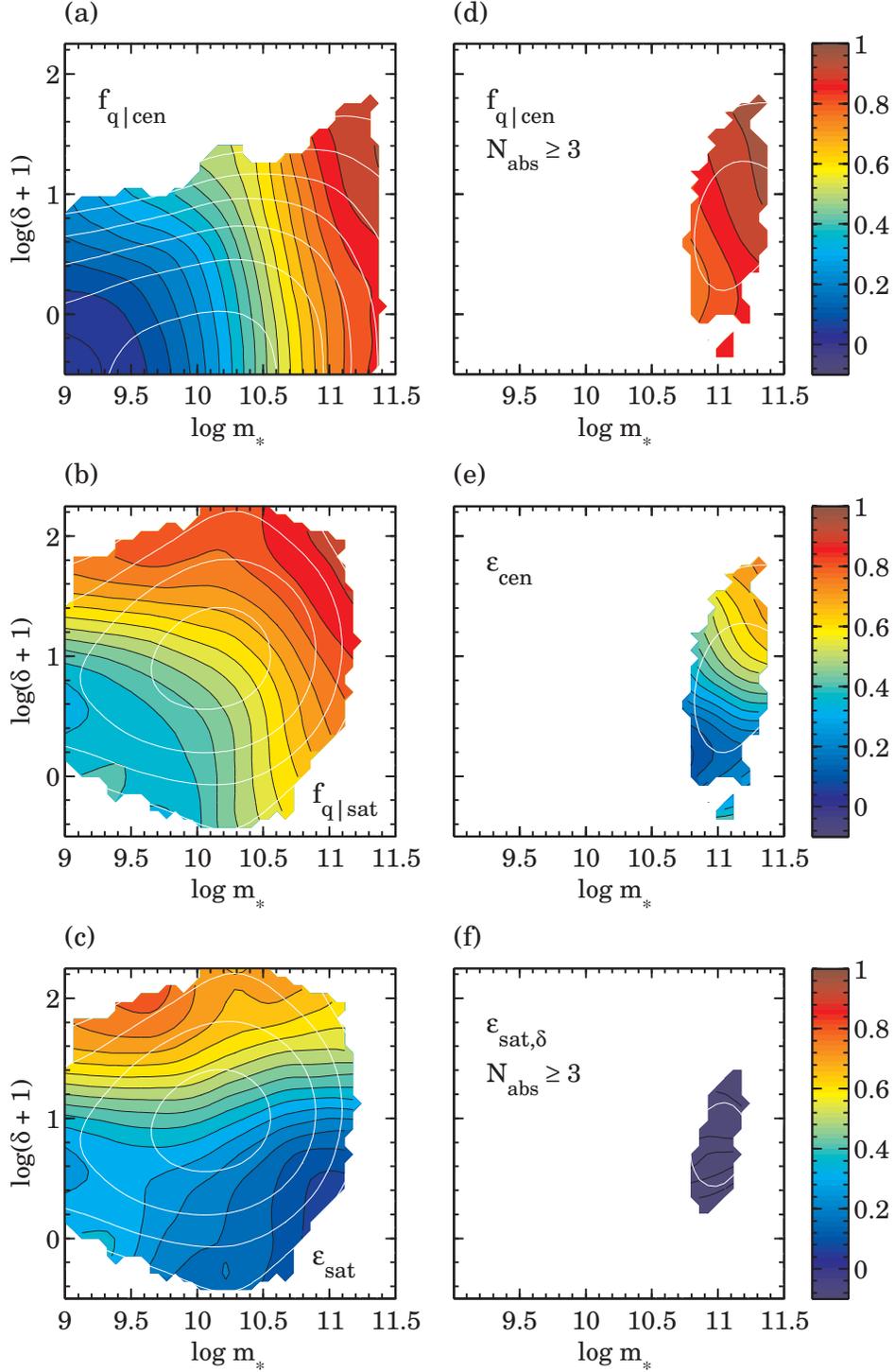}
	\caption{Quenched fractions $f_{\rm q}$ and satellite quenching efficiencies $\epsilon_{\rm sat}$ as a function of stellar mass $m_\ast$ and overdensity $\delta$ for different galaxy samples. Panel (a) shows the quenched fraction for centrals, $f_{\rm q|cen}$, panel (b) the quenched fraction for satellites, $f_{\rm q|sat}$, and panel (c) $\epsilon_{\rm sat}(m_\ast,\delta)$. Panel (d) shows $f_{\rm q|cen}|_{N_{\rm abs} \geq 3}$, i.e., the centrals are only taken from groups, in which our satellites reside (i.e., $N_{\rm abs} \geq 3$). Panel (e) shows $\epsilon_{\rm cen}$ as defined in Equation (\ref{eq:eps_cen}), i.e., the quenching efficiency computed for the centrals in the groups with $N_{\rm abs} \geq 3$, but using the same background $f_{\rm q | cen}$ as used to compute the $\epsilon_{\rm sat}$ in panel (c). Panel (f) shows $\epsilon_{\rm sat,\delta}(m_\ast,\delta)|_{N_{\rm abs} \geq 3}$ (i.e., the satellite quenching efficiency whose background sample of centrals includes the $\delta$ dependence explicitly; see Section \ref{sec:eps_sat}), if the background sample of centrals is restricted to $N_{\rm abs} \geq 3$ (cf.~panel (d)). The white contours indicate the (unweighted) number density of galaxies in logarithmic scale. The color coding is the same for all panels and is indicated by the color bars. To produce the contour plots we have used running bins with size $0.2 \times 0.2$ in logarithmic parameter space and have additionally smoothed the resulting image by a Gaussian filter with standard deviation 0.3. Bins with less than 30 galaxies were discarded.}\label{fig:fq_cen_sat}
\end{figure*}

In the panels (a) and (b) of Figure \ref{fig:fq_cen_sat} we show the quenched fraction of all centrals, $f_{\rm q,cen}(m_{\ast},\delta)$, and of satellites, $f_{\rm q,sat}(m_{\ast},\delta)$, as a function of their stellar mass $m_\ast$ and galaxy overdensity $\delta$. Then in panel (c) we show the corresponding satellite quenching efficiency $\epsilon_{\rm sat}(m_{\ast},\delta)$ computed in the usual way using the mass-dependent $f_{\rm q,cen}(m_{\ast})$.

It is easy to recognize the main features that were earlier found by \cite{peng2012}.  First, the $f_{\rm q | cen}$ at fixed mass of the general central population is indeed only weakly dependent on $\delta$, as illustrated by the near vertical contours over much in panel (a). This is less true at low central masses but the density dependence of $f_{\rm q | cen}$ for these centrals could reflect the misidentification of low mass centrals in high density environments. There are not so many low mass centrals in high density environments and there could be some significant contamination from misidentified satellites (as argued by \citealt{peng2012}).  

The different pattern of $f_{\rm q|sat}$ in panel (b) reflects the combination of mass quenching effects (at high masses) and mass-independent ``environment quenching'' dominating at lower masses.  Once we take out the mass effects via $\epsilon_{\rm sat}$ we also reproduce the remarkably weak dependence of $\epsilon_{\rm sat}$ on $m_\ast$ for fixed $\delta$ in panel (c). Since at low masses, where environmental effects dominate \citep{peng2010}, the number of satellites in high density environments far outweighs the number of centrals, it is clear that, as correctly stated by both \cite{peng2012} and \cite{kovac2014}, satellites are responsible for most of the $\delta$ dependence within the overall population of galaxies.  The dominance of singleton centrals also accounts for the differences in the shape, at low masses, of the mass function of passive central and satellite galaxies that was highlighted in \cite{peng2012}, i.e., the prominence of the second Schechter component for satellites and its absence for centrals.

However, two related points should be noted.  First, as can be seen from the white contours in Figure \ref{fig:fq_cen_sat}, and also from the histograms in Figure \ref{fig:density_field_histogram} and the figures in \cite{peng2012}, the bulk of centrals and satellites occupies different regions in the $(m_\ast,\delta)$ plane, and especially in $\delta$. The peak of the centrals is at $\log(\delta+1) \sim -0.5$, while the peak of the satellites is at $\log(\delta+1) \sim 1$.  Second, 98\% of our centrals inhabit groups with $N_{\rm abs}\leq 2$. That is, the sample of centrals is overwhelmingly dominated by galaxies for which the estimated $\delta$ is based on length scales that are typically much larger than the virial radii of the halos concerned and the weak dependence of $\delta$ for the centrals could conceivably be a consequence of the fact that we are not measuring, for most centrals, the actual local environment, which may be most relevant for the quenching of galaxies. 

To look at the centrals that are in the same groups as the satellites, we first recompute $f_{\rm q | cen}$ after constraining the sample of centrals to groups with $N_{\rm abs} \geq 3$ (see Figure \ref{fig:fq_cen_sat}, panel (d)), i.e., the centrals of the satellites we are considering. Careful inspection shows that the $f_{\rm q | cen}$ of these centrals is systematically higher than for the general set of centrals and is actually comparable or even larger than the $f_{\rm q | sat}$ of the satellites.  To see this more clearly, we show two further panels in Figure \ref{fig:fq_cen_sat}.  

 First, in panel (e) we show a new quenching efficiency that is computed for the centrals in the groups using exactly the same general $f_{\rm q | cen}$ for the background sample as was used to compute the $\epsilon_{\rm sat}$ in panel (c). We denote this new quantity in formal analogy to Equation (\ref{eq:eps_sat_standard}) by
\begin{equation}\label{eq:eps_cen}
\epsilon_{\rm cen}(m_\ast,\delta) = \frac{f_{\rm q | cen}(m_\ast,\delta)|_{N_{\rm abs} \geq 3} - f_{\rm q | cen}(m_\ast)}{1-f_{\rm q | cen}(m_\ast)}\:.
\end{equation}
Here $f_{\rm q | cen}(m_\ast,\delta)|_{N_{\rm abs} \geq 3}$ denotes the quenched fraction of centrals restricted to groups with $N_{\rm abs} \geq 3$. This therefore represents an environmental quenching efficiency for centrals. It can be seen in panel (e) that, over the limited range of mass where this is defined, $\epsilon_{\rm cen}$ is very similar to the $\epsilon_{\rm sat}$ surface, emphasizing the similarity in the $(m_\ast,\delta)$ plane in the quenching of satellites and their associated centrals.  

Second, in panel (f), we show a modified satellite quenching efficiency $\epsilon_{\rm sat,\delta}(m_\ast,\delta)$ that is now computed with the density-dependent $f_{\rm q | cen}$ of the centrals (see Section \ref{sec:eps_sat}) with $N_{\rm abs} \geq 3$.  This is, over the region that can be computed, close to zero or even negative.  This again illustrates the similarity in the quenching of centrals and satellites in the same groups.

\begin{figure}
	\centering
	\includegraphics[width=0.48\textwidth]{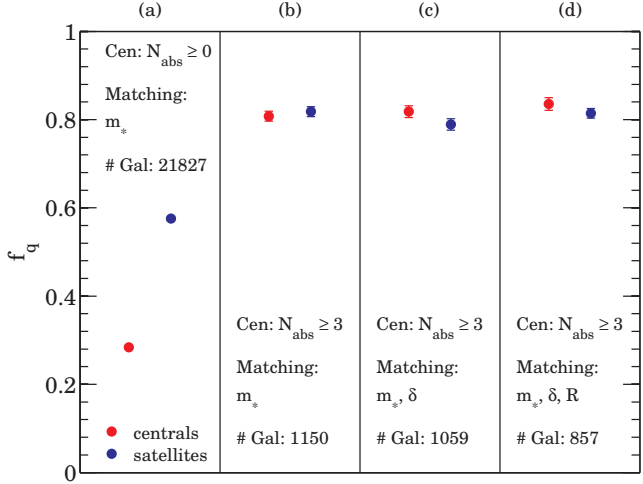}
	\caption{Comparison of the quenched fraction $f_{\rm q}$ of centrals (red) and satellites (blue) for different samples. For panel (a) we used all centrals (i.e., $N_{\rm abs} \geq 0$), while for all other panels we used only the centrals of the groups, in which our satellites are (i.e., $N_{\rm abs} \geq 3$). In the panels (a) and (b) we match the samples of centrals and satellites only with respect to stellar mass, while in panel (c) we match them in $m_\ast$ and $\delta$ and in panel (d) in $m_\ast$, $\delta$, and $R$. The number of galaxies within the matched samples is indicated at the bottom of each panel. If we consider only centrals in groups with $N_{\rm abs} \geq 3$, we do not detect any significant difference in the quenched fraction. This remains true as we additionally match in $\delta$ and $R$.}\label{fig:central_satellite_matched}
\end{figure}

A more direct comparison between $f_{\rm q | cen}$ and $f_{\rm q | sat}$ is given in Figure \ref{fig:central_satellite_matched} that is based on carefully matched samples of satellites and centrals.
Here we compare the mean values of $f_{\rm q | cen}$ and $f_{\rm q | sat}$ for different sets of mass-matched samples of centrals and satellites. In panel (a), we mass-match a set of general centrals (i.e., $N_{\rm abs} \geq 0$) and satellites, without regard to the richness of the groups or to the overdensities $\delta$.  As expected, the mass-matched centrals have a systematically lower $f_{\rm q}$ than the equivalent satellites. This is a reflection of the average $\epsilon_{\rm sat} \sim 0.4$ derived by \cite{vandenbosch2008}, \cite{peng2012}, and at higher redshifts by \cite{knobel2013}, and indeed we can compute from the numbers in this panel an average $\epsilon_{\rm sat}$ for this overall sample of $\epsilon_{\rm sat} \simeq 0.41$.

In the other panels to the right, we now consider only those centrals in $N_{\rm abs} \geq 3$ groups and then additionally match the environment $\delta$ and distance $R$ for the centrals and satellites. It is obvious that the difference between the quenched fractions of centrals and satellites in the leftmost panel completely disappears as soon as we focus on only the $N_{\rm abs} \geq 3$ groups, and remains as we additionally match in $\delta$ and $R$.   However, as can be seen from the numbers in Figure \ref{fig:central_satellite_matched}, the regions in the parameter space where the centrals with $N_{\rm abs} \geq 3$ and satellites overlap are relatively small, so that these carefully matched samples now constitute only a very small fraction of the original samples.  

In summary, we find that we cannot detect any difference in the quenched fractions of centrals and satellites if we focus only on the centrals and satellites in the same $N_{\rm abs} \geq 3$ groups, i.e., if we exclude the vast majority of centrals (i.e., the singletons and pairs), for which $\delta$ is anyway not probing the local density within the halo, and if the centrals and satellites are properly matched in stellar mass and $\delta$. 

This is an important point: in the groups and clusters where the satellites actually reside, the centrals are as quenched as the satellites at a given $(m_\ast,\delta)$, at least in the range of $(m_\ast,\delta)$ that is probed by both centrals and satellites. At least in this restricted part of parameter space, we conclude that centrals do indeed feel environment in the same way as satellites.

We stress that this conclusion does not invalidate the more general statements that were made in \cite{peng2012} but refines them for the particular subset (2\%) of centrals that are in the same halos as the satellites.  It remains to be seen (and we cannot test) whether the environmental effects on massive centrals and satellites at high $\delta$ reflect the same physical effects that are responsible for the environmental effects on lower mass satellites which also dominate the environmental effects in the overall population (e.g., the striking separability in $f_{\rm q}$ in the overall population highlighted by \citealt{peng2010}). However, the mass independence of the $\epsilon_{\rm sat}$ in panel (c) of Figure \ref{fig:fq_cen_sat} suggests that there may indeed be a connection between high and low mass satellites and thus to the centrals in these same environments.  We return to this point in Section \ref{sec:discussion}.

\section{Satellite Quenching Efficiency as a Function of Environment}\label{sec:eps_sat_analysis}

In this section, we will analyze the dependence of the satellite quenching efficiency $\epsilon_{\rm sat}$ on six parameters that describe the satellites and their environments, namely the stellar mass $m_{\rm sat}$ of the satellites, the host halo mass $m_{\rm h}$, the stellar mass of the central $m_{\rm cen}$, the overdensity of galaxies $\delta$, group-centric distance $R$, and the sSFR of the central (see Table \ref{tab:parameters}). To treat the parameters more or less equally, we look at the logarithm of each quantity. Particularly, we measure the sSFR$_{\rm cen}$ offset from the black line in Figure \ref{fig:SSFR_mass_diagram} (in dex) and denote this as $\Delta$sSFR$_{\rm cen}$. The logarithmic widths of the distributions in each parameter are broadly similar, between 1.3 and 3 dex.

\begin{figure*}
   \centering
   \includegraphics[width=0.7\textwidth]{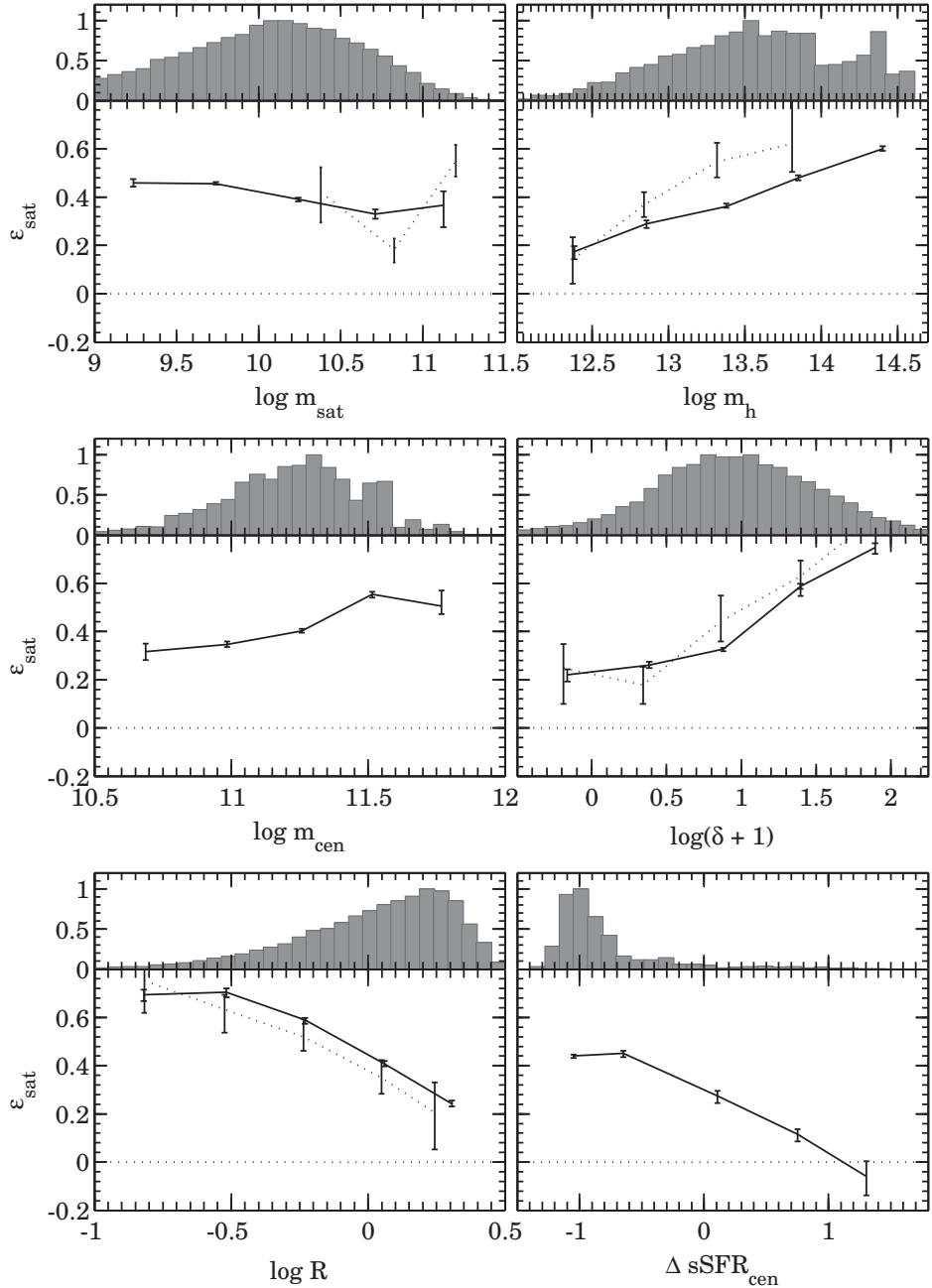}
   \caption{Satellite quenching efficiency $\epsilon_{\rm sat}$ (solid lines) as a function of stellar mass and the environmental parameters (see Table \ref{tab:parameters}). The dotted lines show $\epsilon_{\rm cen}$ (see Equation (\ref{eq:eps_cen}) and cf.~panel (e) of Figure \ref{fig:fq_cen_sat}) indicating that the centrals in the groups are experiencing the same environmental effects as the satellites. For each panel the histogram shows the (unweighted) distribution of the number of satellites within our sample.}\label{fig:e_s_1D}
\end{figure*}

The average satellite quenching efficiency as a function of each of these six parameters individually is shown in Figure \ref{fig:e_s_1D}, where we plot the average $\epsilon_{\rm sat}$ of all satellites in five equally spaced bins of each of the six parameters in turn.  First, the dependence of $\epsilon_{\rm sat}$ on both $m_{\rm sat}$ (and also $m_{\rm cen}$) is very weak. The lack of a significant dependence on the mass of the satellite is a confirmation of the results of \cite{vandenbosch2008} and \cite{peng2012} and we will not discuss it further. On the other hand, we find a strong dependence of the average $\epsilon_{\rm sat}$ on $m_{\rm h}$, $\delta$, and $R$. This is consistent with results previously reported by \cite{peng2012} and \cite{woo2013}.  It is noticeable that $\epsilon_{\rm sat}(\delta)$ seems to change its slope around $\delta \sim 1$ and we will return to this point below. We would expect $\epsilon_{\rm sat}$ to have some dependence on $m_{\rm cen}$ because $m_{\rm cen}$ and $m_{\rm h}$ are coupled.  The link between satellite quenching and the sSFR of the central galaxy has been noted by \cite{weinmann2006} in the context of ``galactic conformity'' and has been analyzed in the $\epsilon_{\rm sat}$ formalism by \cite{phillips2014} at least for galaxy pairs. We devote Section \ref{sec:galaxy_conformity} to an extensive discussion of this sSFR dependence.

As an aside, we further illustrate the similarity of centrals and satellites in the groups discussed in Section \ref{sec:centrals_satellites} by plotting $\epsilon_{\rm cen}$ (see Equation (\ref{eq:eps_cen})) on the four relevant panels of Figure \ref{fig:e_s_1D} as dotted curves. (We choose to represent the dependence of $\epsilon_{\rm cen}$ on the stellar mass of the central on the $m_{\rm sat}$ panel rather than the $m_{\rm cen}$ panel below it, simply because we are interested in the mass of the galaxy itself irrespective of whether it is a central or a satellite.)  It can be seen that these curves are fairly similar, reinforcing our earlier conclusion that the centrals in the groups are experiencing the same environmental effects as the satellites.

\begin{deluxetable*}{llrr}
\tablewidth{0pt}
\tablecaption{The Parameters Used in This Work}
\tablehead{
   \colhead{Parameter} &
   \colhead{Parameter Description} &
\multicolumn{2}{c}{Range} \\
   \colhead{} &
   \colhead{} &
       \colhead{Min}&
       \colhead{Max}  }
\startdata
$\phantom{000}\log m_{\rm sat}$ & Stellar mass of the satellites & $9.0\phantom{0}$ & $11.5\phantom{0}$\\
$\phantom{000}\log m_{\rm h}$ & Halo mass of the group & $12.0\phantom{0}$ & $14.7\phantom{0}$\\
$\phantom{000}\log m_{\rm cen}$ & Stellar mass of the central of the group & $10.5\phantom{0}$ & $12.0\phantom{0}$\\
$\phantom{000}\log(\delta+1)$ & Overdensity based on the fifth nearest neighbor approach (see Section \ref{sec:density_field})& $-0.75$ & $2.25$\\
$\phantom{000}\log R$ & Group-centric distance as defined in Equation (\ref{eq:R}) & $-1.0\phantom{0}$ &  $0.5\phantom{0}$\\
$\phantom{000}\Delta {\rm sSFR}_{\rm cen}$ & Offset (in dex) of log sSFR of the central from the black line in Figure \ref{fig:SSFR_mass_diagram} & $-1.5\phantom{0}$ & $1.8\phantom{0}$
\enddata
\label{tab:parameters}
\end{deluxetable*}

The difficulty in interpreting the one-dimensional plots in Figure \ref{fig:e_s_1D} arises because there are often strong correlations between these six parameters. Table \ref{tab:correlations} gives the correlation coefficients between the five parameters (excluding the sSFR of the central). If two parameters $p_1$ and $p_2$ are correlated and $p_1$ is driving the slope of the satellite quenching efficiency, the correlation between $p_1$ and $p_2$ could induce a correlation between $p_2$ and $\epsilon_{\rm sat}$ which is of no physical relevance.  Looking at Table \ref{tab:correlations}, we observe that, on the one hand, there is a strong correlation between $\log m_{\rm h}$ and $\log m_{\rm cen}$ for the satellites and, on the other hand, there is a strong correlation (or anti-correlation) between $\log m_{\rm h}$, $\log(\delta+1)$, and $\log R$. To unravel these correlations and to try to gain more information on the physical origins of $\epsilon_{\rm sat}$, it is useful to move from the one-dimensional plots of Figure \ref{fig:e_s_1D} to two-dimensional plots for pairs of the parameters displaying the strongest one-dimensional effects.

\begin{figure}
   \centering
   \includegraphics[width=0.48\textwidth]{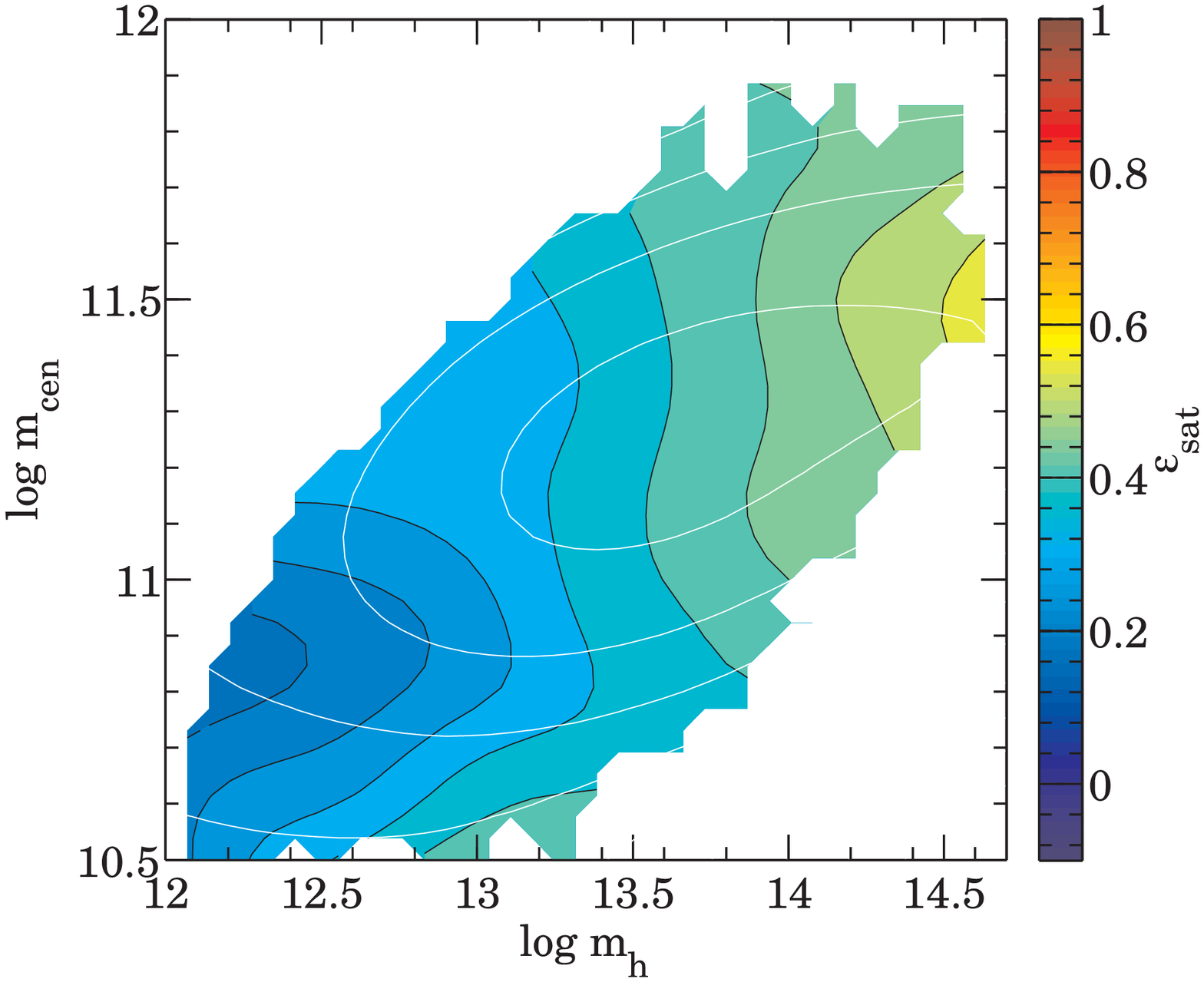}
   \caption{Satellite quenching efficiency $\epsilon_{\rm sat}$ as a function of halo mass $m_{\rm h}$ and stellar mass of the central $m_{\rm cen}$. The colors correspond to the values of $\epsilon_{\rm sat}$ as indicated in the color bar. The white contours indicate the (unweighted) number density of galaxies in logarithmic scale. To produce the contour plots we have used running bins with size $0.2 \times 0.2$ in logarithmic parameter space and have additionally smoothed the resulting image by a Gaussian filter with standard deviation 0.3. Bins with less than 30 galaxies were discarded.}\label{fig:fq_Mh_Mc}
\end{figure}

We look first at Figure \ref{fig:fq_Mh_Mc}, which shows $\epsilon_{\rm sat}(m_{\rm h},m_{\rm cen})$. The gradient of $\epsilon_{\rm sat}$ is basically pointing entirely in the direction of $m_{\rm h}$ for $\log m_{\rm h} > 13.5$ (i.e., the contours are more or less vertical).  For $\log m_{\rm h} < 13.5$ the pattern of the contours is more complex, but still there is no clear trend with $m_{\rm cen}$ visible. It is not clear, whether the deviations from vertical lines reflect an actual signal or whether they are due to misclassifications of centrals in low mass groups and/or misidentification of low mass groups in general. We can conclude that at fixed halo mass, the stellar mass of the central galaxy seems to have no influence of the probability that a satellite is quenched beyond the effect of the halo mass (at least for $\log m_{\rm h} > 13.5$).

\begin{deluxetable}{l|ccccc}
\tablewidth{0pt}
\tablecaption{Pearson Correlation Coefficients Between the Environment Parameters for the Satellites in Our Sample}
\tablehead{
\colhead{} &
   \colhead{$\log m_{\rm sat}$} &
       \colhead{$\log m_{\rm h}$} &
   \colhead{$\log m_{\rm cen}$} &
   \colhead{$\log(1+\delta)$} &
   \colhead{$\log R$}
 }
\startdata
$\log m_{\rm sat}$ &  $\phantom{-}1.00$ & $\phantom{-}0.04$ &   $\phantom{-}0.06$  & $\phantom{-}0.14$  & $-0.08$ \\
$\log m_{\rm h}$ &   $\phantom{-}0.04$ & $\phantom{-}1.00$   & $\phantom{-}0.69$ & $\phantom{-}0.41$  & $-0.08$ \\
$\log m_{\rm cen}$ &  $\phantom{-}0.06$ & $\phantom{-}0.69$   & $\phantom{-}1.00$ & $\phantom{-}0.26$  & $-0.04$ \\
$\log(1+\delta)$  &  $\phantom{-}0.14$ & $\phantom{-}0.41$  & $\phantom{-}0.26$ & $\phantom{-}1.00$  & $-0.52$ \\
$\log R$ &  $-0.08$&   $-0.08$ & $-0.04$ &  $-0.52$ &   $\phantom{-}1.00$
\enddata
\tablecomments{For the definition of the parameters see Table \ref{tab:parameters}.}
\label{tab:correlations}
\end{deluxetable}

\begin{figure}
   \centering
   \includegraphics[width=0.48\textwidth]{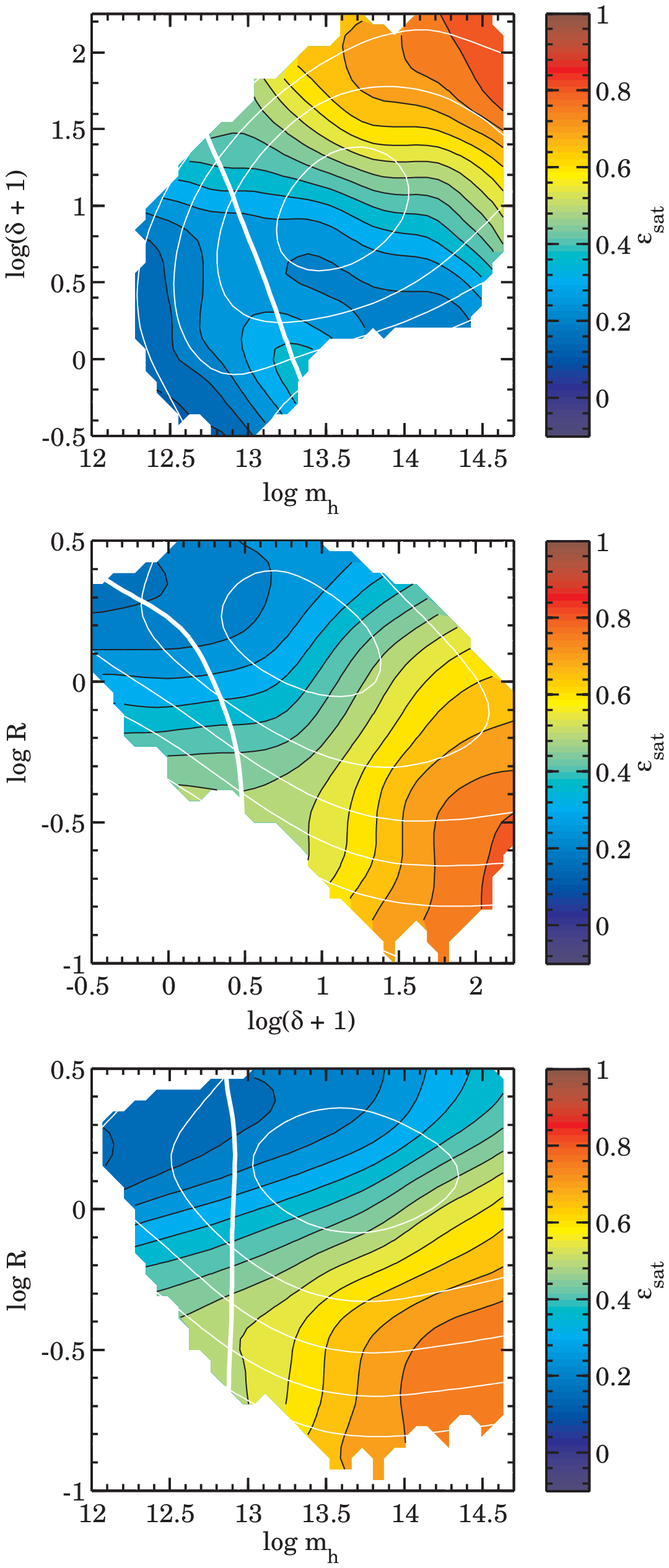}
   \caption{Satellite quenching efficiency $\epsilon_{\rm sat}$ as a function of the pairwise parameters $m_{\rm h}$, $\delta$, and $R$. The colors correspond to the values of $\epsilon_{\rm sat}$ as indicated in the color bar. The thin white contours indicate the (unweighted) number density of galaxies in logarithmic scale and the thick white contour indicates the contour for which the median richness of the satellites is $N_{\rm abs} = 5$. To produce the contour plots we have used running bins with size $0.2 \times 0.2$ in logarithmic parameter space and have additionally smoothed the resulting image by a Gaussian filter with standard deviation 0.3. Bins with less than 30 galaxies were discarded.}\label{fig:Mh_delta_r}
\end{figure}

More interesting are the two-dimensional plots between each of the three parameters which have relatively strong effects in the one-dimensional plots, i.e., $\log m_{\rm h}$, $\log(\delta+1)$, and $\log R$. These are shown in Figure \ref{fig:Mh_delta_r}.  Given the issues discussed earlier in the interpretation of the overdensity $\delta$, the thick white line in each panel indicates the contour for which the median richness of the satellites is $N_{\rm abs} = 5$, i.e., it separates the plane into regions that are dominated by rich groups, for which $\delta$ is measuring the local density, and small groups, for which $\delta$ is less informative. We can immediately observe the following trends.

On the right side of the thick white line in the top panel (i.e., in rich groups), the gradient of $\epsilon_{\rm sat}$ has a strong component in the direction of $\delta$. In fact, the gradient is stronger in the direction of $\delta$ than in the direction of $m_{\rm h}$. However, on the left side of the thick white line, the dependence on $\delta$ is almost negligible. This behavior largely explains the origin for the change of slope of $\epsilon_{\rm sat}(\delta)$ in Figure \ref{fig:e_s_1D}.  A primacy of $\delta$ over $m_{\rm h}$ in the regime where $\delta$ is most meaningful was pointed out by \cite{peng2012}. On the right side of the thick white line in the middle panel, the dependence of $\epsilon_{\rm sat}$ on the group-centric distance $R$ is similarly important as for $\delta$. On the left side, where $\delta$ is less meaningful, however, $R$ is the dominating parameter (as was also found by \citealt{carollo2014}).  However, on all three panels of Figure \ref{fig:Mh_delta_r} the contours are not simply horizontal or vertical, and so we must conclude that all of $\delta$, $R$ and $m_{\rm h}$ are playing a role in driving quenching, with relative importance depending on the region of parameter space in consideration.

This conclusion is further confirmed by a more quantitative analysis in the five-dimensional parameter space including $m_\ast$, $m_{\rm h}$, $m_{\rm cen}$, $\delta$, and $R$. To try to disentangle the correlations between different parameters, we identified pairs of galaxies aligned along a given dimension (i.e., which differed primarily in only one of the parameters). From all these pairs aligned along one dimension, we estimated an effective mean gradient along this dimension. We find that these gradients (normalized by the width of the parameters space given in Table \ref{tab:parameters}) for $m_{\rm h}$, $m_{\rm cen}$, and $\delta$ are all of similar strength and inconsistent with zero at about $3\sigma$. On the other hand, the effective gradients for both $m_\ast$ and $m_{\rm h}$ are consistent with zero.

Here, we briefly comment what we mean by saying ``parameter $p_1$ in a certain regime more dominant than parameter $p_2$''. We mean that the gradient of $\epsilon_{\rm sat}(p_1,p_2)$ normalized to the typical range of relevant values of $p_1$ and $p_2$ in log scale has a larger component in the direction $p_1$ than $p_2$. Since the axes in Figure \ref{fig:Mh_delta_r} roughly correspond to these ranges, the dominant parameter is the one in whose direction most contour lines are crossed. However, due to the rather vague definition of ``parameter space'' and due to the fact that the gradient of $\epsilon_{\rm sat}$ is changing over the parameter space, it is difficult to precisely quantify this concept.  A further difficulty arises because the measurement uncertainty in each of these three parameters almost certainly varies across the diagrams.

Most of these conclusions have been seen before in similar analyses, which have however been more limited or have not used the quenching efficiency formalism.  One of the goals of this section has been to present these results in a single coherent and homogeneous way.

To summarize, we conclude from this analysis that all of $\delta$, $R$, and $m_{\rm h}$ are playing a role in the environmentally driven quenching of satellites.  The stellar mass of the satellite and of the central are not correlated with $\epsilon_{\rm sat}$, except to the extent that the latter may reflect halo mass.  Finally, the sSFR of the central is a major factor, and it is to this effect that we now turn in the next section.

\section{Galactic Conformity}\label{sec:galaxy_conformity}

We showed in Figure \ref{fig:e_s_1D} that one of the strong environmental dependencies of $\epsilon_{\rm sat}$ is the sSFR of the central, which is parameterized as $\Delta {\rm sSFR}_{\rm cen}$ (see Table \ref{tab:parameters}) being the offset in sSFR from the black line in Figure \ref{eq:SSFR_division}. In this section, we will analyze and discuss this effect in detail. Since the distribution of satellites strongly peaks around $\Delta {\rm sSFR}_{\rm cen} \simeq -1$ (i.e., $92\%$ of the satellites have a central that is quenched), we will only consider for our analysis, whether $\Delta {\rm sSFR}_{\rm cen}$ for a given satellite is larger or smaller than zero (i.e, whether the central of a given satellite is star forming or quenched).

\subsection {The Meaning of Galactic Conformity}\label{sec:meaning_conformity}

Since it was introduced by \cite{weinmann2006}, the term ``galactic conformity'' has entered widespread use in the lexicon of galaxy evolution to describe the effect whereby quenched centrals are more likely to have quenched satellites (and vice versa).  We think it is important to clarify the use and astrophysical implications of this word.

We first distinguish between the probability that a particular galaxy is quenched, $P_{\rm q}$, and the observed fraction of similar galaxies that have actually been quenched, i.e., the outcome $f_{\rm q}$ that we have used previously in this paper.  Generally, we would regard the latter as a good estimator of the former, neglecting complications such that $P_{\rm q}$ may have varied with time and $f_{\rm q}$ is usually the integrated outcome observed at one particular epoch.

However, galaxies are (probably) not intrinsically probabilistic systems, and therefore the probabilistic aspect of $P_{\rm q}$ must reflect the action of some number of ``hidden variables'' --- to use the parlance of quantum mechanics --- in the general sense of ``missing information'' which would be necessary to completely understand the causal physical process of galaxy quenching. Some of these variables may be known and their effect can, if desired, be removed by considering $P_{\rm q}$ as a function of those variables, i.e., $P_{\rm q}(p_1,p_2,\ldots)$ as estimated by $f_{\rm q}(p_1,p_2,\ldots)$.  The action of all other relevant variables that are not explicitly treated in this way will be to introduce a ``probabilistic'' aspect to quenching.  Some of the relevant variables may well be unknown or impractical to observe (or even reflect some complicated physical process that cannot be adequately described with a few global parameters).  However, if we had complete knowledge of all the relevant parameters involved in quenching, then quenching would (presumably) no longer be probabilistic.

The term ``galactic conformity'' was introduced by \cite{weinmann2006} to describe the situation where there is a correlation between the state of the centrals and that of the satellites (i.e., that quenched centrals generally have quenched satellites and star-forming centrals have star-forming satellites), {\it even when the two samples of satellites have been carefully matched in one or more parameters}.  In other words, the $f_{\rm q}(p_1,p_2,\ldots)$ for the satellites of quenched centrals is higher than the $f_{\rm q}(p_1,p_2,\ldots)$ of satellites of non-quenched centrals at the same values of $p_1,p_2,\ldots$. In the particular case of \cite{weinmann2006}, the parameters that were matched were the halo mass and the luminosity of the satellites.

It is clear that an overall association of quenched centrals and satellites will arise rather trivially, if the individual probabilities that centrals and satellites be quenched, i.e., $P_{\rm q,cen}$ and $P_{\rm q,sat}$, respectively, both depend on some external parameter whose value is shared by both the centrals and satellites. One could imagine, as did \cite{weinmann2006}, that the chance for galaxies to be quenched depends on the mass of the parent DM halo, which will be the same for the central and all the satellites in a given halo.   In this case, low mass halos would statistically have a lower quenched fraction of both centrals and satellites than would high mass halos, and therefore quenched centrals would generally be found with quenched satellites \citep[cf.][]{wang2012}.  If the effect of the halo mass was extremely strong, i.e., if there was a sharp threshold in halo mass above which all galaxies were quenched, then the resulting association of quenched satellites with quenched centrals would likewise be very strong, in that essentially all quenched centrals would have quenched satellites, and vice versa.

However, in the simple situation described in the previous paragraph, if we now consider just a single value of halo mass or, equivalently, if we carefully match the distribution of halo masses of the satellites with quenched and non-quenched centrals, then we would find that the preferential association of quenched centrals and satellites would vanish. The distribution of halo masses of the satellites with quenched and non-quenched centrals would by construction be exactly the same, and therefore, if the quenching probability depended only on the halo mass, there would be no difference in the fraction of quenched satellites for the two sets of centrals, i.e., no ``conformity''. Conformity would have been seen in a general set of centrals and satellites (as described in the previous paragraph), but would then disappear once we matched the samples of satellites in halo mass.  The existence or non-existence of galactic conformity therefore depends critically on the construction and analysis of the galaxy sample and particularly on the choice of any parameters that have been matched between the samples of satellites with quenched and non-quenched centrals.

The meaning of galactic conformity (in a given sample) is that the satellites somehow ``know'' the outcome of whether the central was quenched (and vice versa).  This is equivalent to saying that there must be one or more ``hidden variables'' that are influencing the quenching of both centrals and satellites, but are not being matched in the sample in question.  Furthermore, these hidden variable(s) must also be correlated (or even the same) for the centrals and satellites in a given halo.  If both these hold, then the values of these ``hidden common variable(s)'' will be different for the quenched and non-quenched samples of galaxies, leading to the correlation of quenched satellites with quenched centrals that we call galactic conformity, even when other known variables have been matched.

Given the extensive correlations between the variables that are encountered in descriptions of galaxies and their environments, a hidden variable could well be correlated with one of the variables that had previously been matched.  However, in this case, it is clear that it is only that component of the hidden variable that is orthogonal to those previously matched variables which can cause any (remaining) conformity.

If we were now to include a previously hidden variable (that was previously producing conformity) by additionally matching the values of this variable between the samples of satellites with quenched and non-quenched centrals, then the conformity would disappear (or at least be reduced if further hidden variables remained hidden). We stress that conformity should therefore be thought of as arising from the analysis rather than being some kind of absolute physical effect. The same set of galaxies will show conformity when analyzed in one way, but this will be reduced or eliminated as more variables are introduced and matched.

Obvious possibilities for potential hidden common variables include variables that are (1) relatively easy to observe and therefore ``unhide'', like halo mass or local overdensity (as in this paper), or (2) presently unobservable for most groups, like the gas entropy or the presence of shock heating or of cold gas flows, or (3) almost unobservable, such as the time that has elapsed since some particular event such as group formation.

As an example of the last case, \cite{hearin2014} argued that the ``two-halo conformity'' (i.e., conformity on scales larger than the virial radius, which we have not studied in this paper) could arise naturally as a consequence of an ``assembly bias''. To demonstrate this, they produced simulated galaxy mock catalogs for SDSS, in which --- at a given stellar mass --- SFRs were randomly drawn from the corresponding distribution of SDSS galaxies (at this mass) and assigned to the mock galaxies (at this mass) according to the relative ``age'' of their (sub)halos (i.e., the older the (sub)halo the lower the SFRs). The ``age'' of a (sub)halo was quantified by the (sub)halo property $z_{\rm starve}$, which correlates with the epoch of star formation cessation within the (sub)halo \citep[for details, see][]{hearin2013}. They did not differentiate between centrals and satellites nor did they attribute any special role to the virial radius of the halos. As a result they observe a strong conformity signal within these mock catalogs on the scale 1-5 Mpc without the need of introducing any additional quenching for the satellites after they fall into the halos.

Another important complication is that the (unknown) residuals that arise from observational scatter in the practical estimation of the observables (like halo mass) will effectively also act as a hidden variable. 

To summarize, the existence of galactic conformity in a given sample and in a given analysis tells us primarily that there are still additional hidden variables which must affect in some way the quenching of both central and satellite galaxies and whose value must be shared in some way by centrals and satellites.  When a complete description of quenching (with noiseless data) is achieved, then conformity will have disappeared.

\subsection{Galactic Conformity in the Current Sample}\label{sec:current_conformity}

In the following, we denote the satellite quenching efficiency for satellites with a star-forming central by $\epsilon_{\rm sat,SF}$ and for satellites with a quenched central by $\epsilon_{\rm sat,q}$. Averaged over all parameters, we measure a mean satellite quenching efficiency $\langle \epsilon_{\rm sat,q} \rangle \simeq 0.44$ around quenched centrals and $\langle \epsilon_{\rm sat,SF} \rangle \simeq 0.17$ for the $\sim \! 8\%$ of satellites that are found around star-forming centrals.  In other words, within the general (unmatched) satellite population and taking out the effects of stellar mass (by using $\epsilon_{\rm sat}$), the environmentally driven quenching of satellites is $2.6$ times stronger in satellites with quenched centrals than in those with star-forming centrals.  However, in the context of the discussion in Section \ref{sec:meaning_conformity}, this difference could arise simply because of the dependencies of $\epsilon_{\rm sat}$ on the other variables (i.e., $m_{\rm sat}$, $m_{\rm h}$, $m_{\rm cen}$, $\delta$, $R$) that is shown in Figure \ref{fig:e_s_1D}, if these variables are also affecting the quenching of the centrals.

\begin{figure}
   \centering
\includegraphics[width=0.48\textwidth]{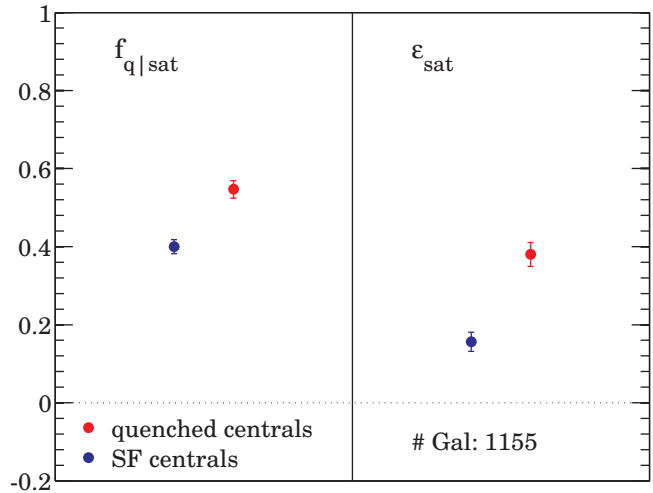}
   \caption{Quenched fraction $f_{\rm q}$ and satellite quenching efficiency $\epsilon_{\rm sat}$ for two matched samples of satellites. The red and the blue points correspond to satellites with quenched centrals and to satellites with star-forming centrals, respectively, which have been matched to each other with respect to stellar mass $m_{\rm sat}$ and all four remaining environmental parameters (i.e., $m_{\rm h}$, $m_{\rm cen}$, $\delta$, and $R$). The number of galaxies within the matched samples is indicated in the right panel. For the mean parameter values of the matched sample see Table \ref{tab:sample_mean}.}\label{fig:galaxy_conformity_matched}
\end{figure}

Accordingly, we construct a sample of satellites with quenched centrals that is matched to the sample of satellites with non-quenched centrals with respect to {\it all} five of these parameters.  The result is shown in Figure \ref{fig:galaxy_conformity_matched}.
We can see immediately that, although the mean $\langle \epsilon_{\rm sat,q}\rangle \simeq 0.38$ of the matched sample and the mean $\langle \epsilon_{\rm sat,SF} \rangle \simeq 0.16$ of the matched sample are both slightly lower, there is still a clear difference between the two satellite quenching efficiencies.   

\begin{deluxetable*}{l|ccccc}
\tablewidth{0pt}
\tablecaption{Weighted Mean Values and Standard Deviations for the Matched and Unmatched Sample of Satellites (see Section \ref{sec:current_conformity} and Figure \ref{fig:galaxy_conformity_matched})}
\tablehead{
\colhead{Sample} &
   \colhead{$\langle \log m_{\rm sat} \rangle$} &
   \colhead{$\langle \log m_{\rm h} \rangle$} &
       \colhead{$\langle \log m_{\rm cen} \rangle$} &
   \colhead{$\langle \log(\delta + 1) \rangle$} &
   \colhead{$\langle \log R \rangle $} 
 }
\startdata
Total sample & $9.8 \pm 0.6$ & $13.6 \pm 0.6$ & $11.2 \pm 0.3$ & $1.0 \pm 0.5$ &  $0.0 \pm 0.3$ \\
Matched sample & $9.8 \pm 0.6$ & $13.2 \pm 0.5$ & $11.0 \pm 0.2$ & $0.8 \pm 0.5$ &  $0.1 \pm 0.2$ 
\enddata
\tablecomments{For the definition of the parameters see Table \ref{tab:parameters}.}
\label{tab:sample_mean}
\end{deluxetable*}

In other words, after removing the effects of stellar mass and the four environmental variables through this matching of the samples, we still find that satellites around quenched centrals are $2.4$ times more likely to be environmentally quenched than those around non-quenched centrals, very similar to the factor of 2.6 obtained simply by averaging across all satellites without any matching. It should be noted that a consequence of our matching is that we are confined to a reduced range of the overall parameter space and so these numbers are not directly comparable.  The (weighted) mean values and (weighted) standard deviations of the five environmental parameters in the matched sample of satellites are compared with the values for the overall, unmatched population of satellites in Table \ref{tab:sample_mean}. It is mainly the halo mass that is different.

As discussed in Section \ref{sec:meaning_conformity}, the persistence of conformity even after these five variables are matched implies that there must still be additional ``hidden'' variables at play that (1) affect the quenching of both satellites and centrals and (2) are distributed across the halo in the sense that satellites and centrals must share, at least to some degree, the values of these variables.

We also noted in Section \ref{sec:meaning_conformity} that the conformity can be produced from residual observational errors in common variables that are playing a role in quenching (such as halo mass).  However, we see no increase in conformity when we artificially introduce additional noise into these parameters and so we are quite confident that such errors represent a negligible contribution to the strong conformity effects presented in this section.

\subsection{Variations in the Strength of Conformity across the Parameter Space}

In order to try to gain clues as to the nature of the missing variable(s), we can look at the strength of the conformity effect as functions of each of the identified variables in turn.   In other words, we can simply split the average $\epsilon_{\rm sat}$ that was plotted against each of the five variables in Figure \ref{fig:e_s_1D} into the contributions of those satellites with quenched and with non-quenched centrals, i.e., $\epsilon_{\rm sat,q}$ and $\epsilon_{\rm sat,SF}$, respectively.

\begin{figure*}
   \centering
\includegraphics[width=0.7\textwidth]{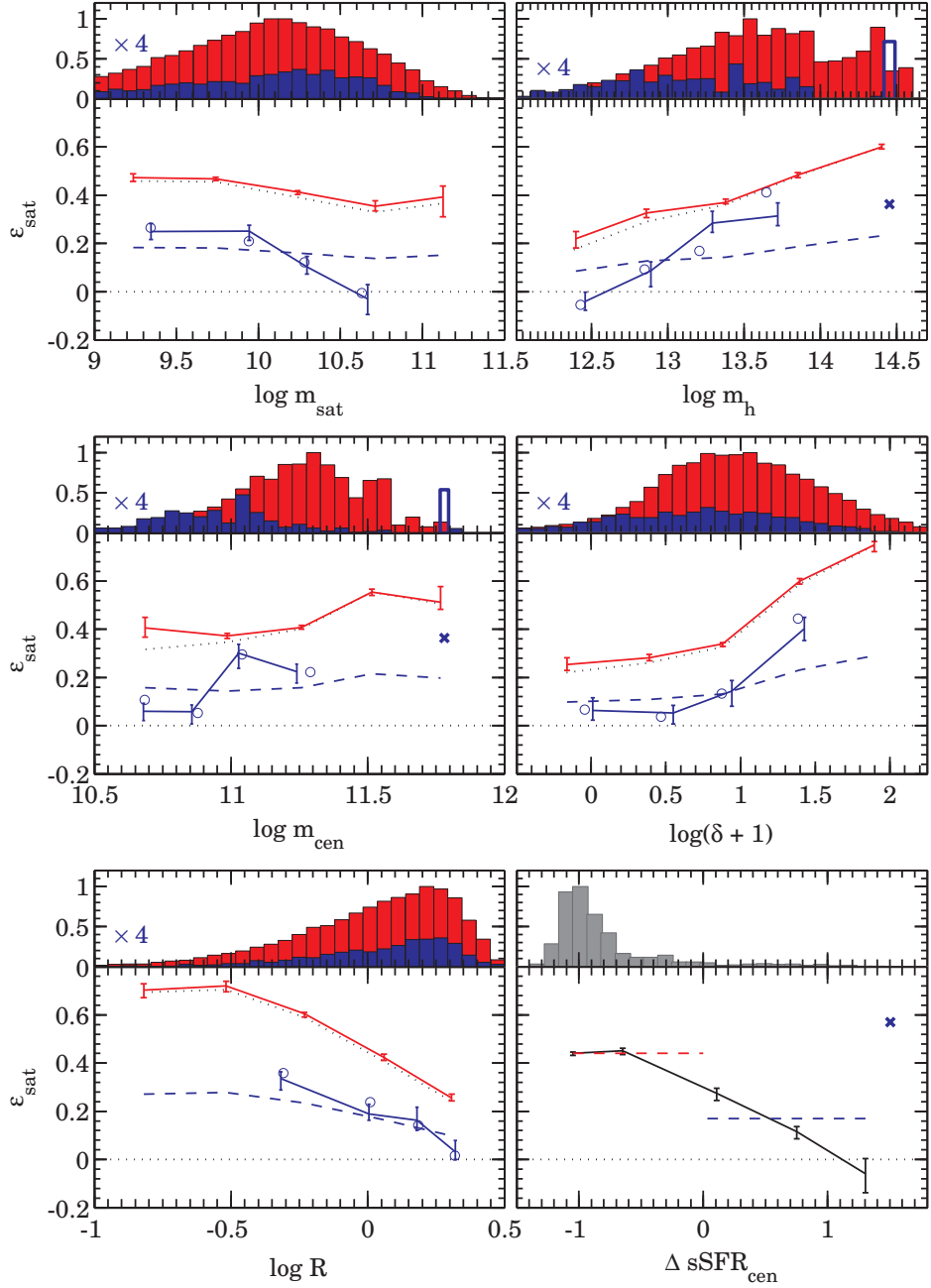}
   \caption{Satellite quenching efficiency $\epsilon_{\rm sat}$ as a function of stellar mass and the environmental parameters (see Table \ref{tab:parameters}) for satellites which have star-forming centrals (i.e., $\epsilon_{\rm sat,SF}$; blue solid lines) and for satellites which have quenched centrals (i.e., $\epsilon_{\rm sat,q}$; red solid lines). The blue open circles correspond to $\epsilon_{\rm sat,SF}$ for the high purity sample of satellites (see Section \ref{sec:group_catalog}). Except for the lowermost right panel, the blue dashed lines correspond to the red lines that have been down scaled by $\langle \epsilon_{\rm sat,SF} \rangle / \langle \epsilon_{\rm sat,q} \rangle \simeq 0.39$ (i.e., it corresponds to the fiducial case, where galactic conformity is constant as a function of our parameters). For each of these panels, the red histogram shows the (unweighted) distribution of satellites that have a quenched central and the blue histograms the (unweighted) distribution of satellites that have a star-forming central. The blue histogram has been artificially enhanced by a factor of four relative to the red histogram for better visibility. As a benchmark, $\epsilon_{\rm sat}$ for all satellite galaxies (see Figure \ref{fig:e_s_1D}) is indicated as a black dotted line. In the lowermost right panel, the black curve and the gray histogram show $\epsilon_{\rm sat}$ and the corresponding distribution, respectively, for all satellites (see Figure \ref{fig:e_s_1D}), whereas the blue and the red dashed lines simply correspond to $\langle \epsilon_{\rm sat,SF} \rangle$ and $\langle \epsilon_{\rm sat,q} \rangle$, respectively. The blue crosses and the blue unfilled histograms correspond to two particular, big clusters with star-forming centrals, which were excluded from the sample (see Section \ref{sec:group_catalog}).} \label{fig:e_s_1D_conformity}
\end{figure*}

These are shown as the red and the blue solid lines, respectively, in Figure \ref{fig:e_s_1D_conformity}.  In each panel, the black dotted line is the overall average $\epsilon_{\rm sat}$ transferred from Figure \ref{fig:e_s_1D}.  Given that the vast majority of satellites have quenched centrals, it is not surprising that this is very similar to the red line for the satellites of quenched centrals.  The blue dashed line is what would be obtained for the satellites of non-quenched centrals if the $\epsilon_{\rm sat,q}$ is uniformly decreased by the factor of $\sim \! 2.5$ that was derived above as the ratio of the average values of $\epsilon_{\rm sat,q}$ and $\epsilon_{\rm sat,SF}$.   If the observed blue solid line lies on this blue dashed line, then it implies that the strength of the conformity is ``average''. If the blue solid line lies below the blue dashed line, i.e., with $\epsilon_{\rm sat,SF}$ approaching zero, it indicates a strong conformity effect at this position in the parameter space in the sense that the satellites of star-forming centrals are hardly being quenched at all by environmental processes.  Lines lying above, with $\epsilon_{\rm sat,SF}$ approaching $\epsilon_{\rm sat,q}$, indicate that there is a weak conformity effect at this location, as the difference between the satellites of quenched and non-quenched centrals is becoming small.  The data are quite noisy, because so few satellites have star forming centrals and so these plots should not be over-interpreted.

In order to check the robustness of our result, we also plot $\epsilon_{\rm sat,SF}$ for the high purity sample of satellites (blue open circles), which was defined in Section \ref{sec:group_catalog}. $\epsilon_{\rm sat,SF}$ for this sample is very similar to $\epsilon_{\rm sat,SF}$ for the general sample of satellites (blue solid lines) and almost always within the error bars of the blue lines. This demonstrates that our conformity signal is, if at all, only weakly affected by uncertainties due to possible misclassification of centrals.

As an aside, it should also be noted in the context of the discussion in Section \ref{sec:meaning_conformity} that the conformity that is evident on the different panels of Figure \ref{fig:e_s_1D_conformity} as the difference between the red and blue solid lines is the conformity that is associated with matching just the one parameter that is plotted as the horizontal axis in each panel.  In other words the conformity seen in the upmost right panel is the conformity that remains once the halo mass (alone) is matched, as in \cite{weinmann2006}, and so on.

The strongest variation in conformity strength in Figure \ref{fig:e_s_1D_conformity} is with halo mass $m_{\rm h}$, where we see a very strong conformity effect at $\log m_{\rm h} \sim 12.5$ (where there are essentially no environmentally quenched satellites around star forming centrals) and a negligible conformity effect at $\log m_{\rm h} \sim 13.5$ (where the satellite quenching efficiency is $\epsilon_{\rm sat} \sim 0.4$, almost independent of the state of the central).

Interestingly, we see little evidence in Figure \ref{fig:e_s_1D_conformity} that the strength of conformity changes with group-centric distance. If anything, conformity appears to be weaker at small $R$, whereas one might have expected to have seen the effect of the central most strongly imprinted on the satellites at small radii.

We can try to quantify the effect, or strength, of conformity in a number of ways.  One way would be to consider $\epsilon_{\rm sat,SF}$ and then consider an additional ``conformity quenching efficiency'' $\epsilon_{\rm conf}$ as an additional quenching term that acts on the surviving star forming satellites when the central becomes quenched.  It is then easy to show that 
\begin{equation}
\epsilon_{\rm conf} = \frac{\epsilon_{\rm sat,q}-\epsilon_{\rm sat,SF}}{1-\epsilon_{\rm sat,SF}} \:.
\end{equation}
However, it may be more natural to consider the effect of conformity as a boosting of the quenching effect of the environment if the central is quenched, or equivalently, as a suppression of the effect of the environment if the central is still star forming.  We may represent this suppression factor $\xi_{\rm conf}$ as
\begin{equation}
\xi_{\rm conf} = \frac{\epsilon_{\rm sat,SF}}{\epsilon_{\rm sat,q}}\:,
\end{equation}
which is simply the inverse of the corresponding boost factor. Conformity is therefore strong when $\xi_{\rm conf}$ is small, and weak when $\xi_{\rm conf}$ approaches unity. The dashed blue lines in Figure \ref{fig:e_s_1D_conformity} would therefore represent an average strength of conformity, $\xi_{\rm conf} \simeq 0.39$.   We choose to consider the suppression $\xi_{\rm conf}$, rather than the boost $\xi_{\rm conf}^{-1}$, simply to have a parameter that varies between zero and unity, and to have $\epsilon_{\rm sat,SF}$ which is generally small and less accurately known than $\epsilon_{\rm sat,q}$ as the numerator of the ratio.  It is a curious coincidence that the overall average value of $\xi_{\rm conf} \simeq 0.39$ is very similar to the overall average value of $\epsilon_{\rm sat} \sim 0.4$ \citep[][and this work]{vandenbosch2008,peng2012}, but two quantities are physically quite distinct, and this coincidence is just that.

It is important to appreciate that the strength of conformity could vary for several reasons.  In terms of our previous discussion in Section \ref{sec:meaning_conformity}, it means that the contribution of the particular ``hidden common parameter(s)'' is varying.   

\begin{figure}
	\centering
	\includegraphics[width=0.47\textwidth]{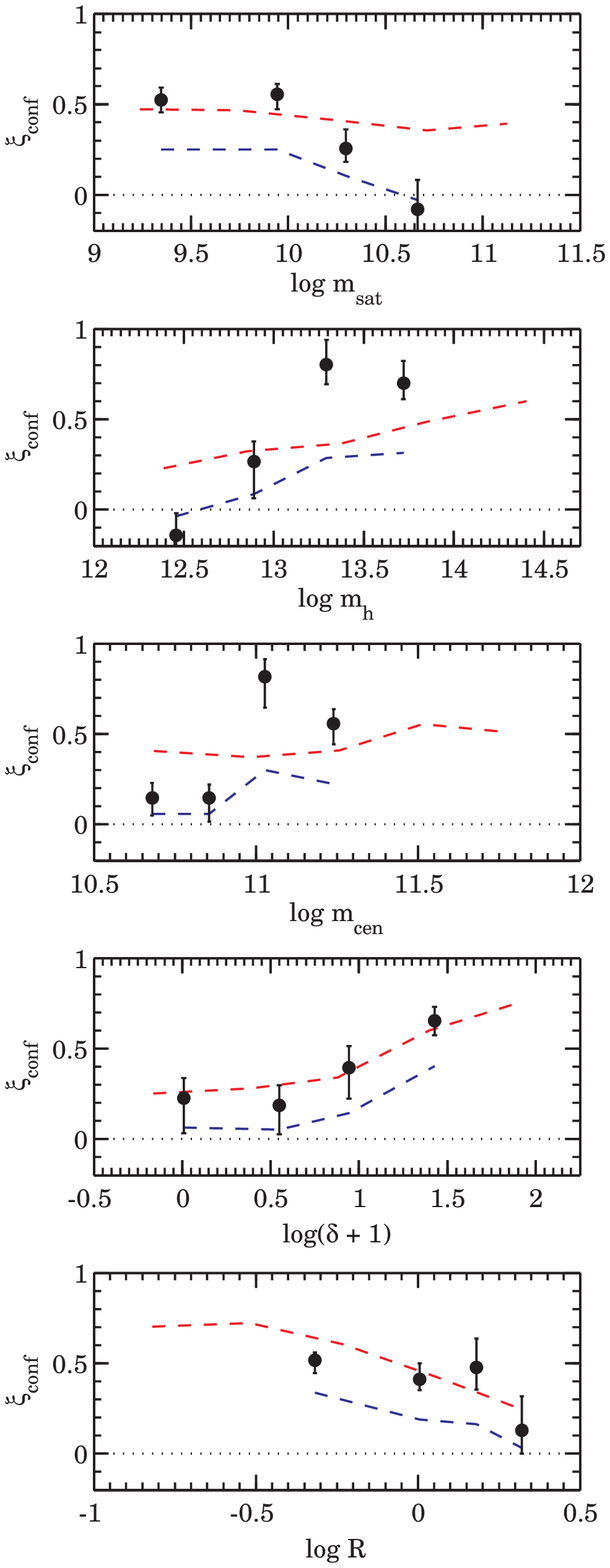}
	\caption{Conformity strength $\xi_{\rm conf} = \epsilon_{\rm sat,SF} / \epsilon_{\rm sat,q}$ as a function of stellar mass and the environmental parameters. The red and blue dashed lines correspond to $\epsilon_{\rm sat,SF}$ and $\epsilon_{\rm sat,q}$, respectively (see Figure \ref{fig:e_s_1D_conformity}). There is a large variation of the values of $\xi_{\rm conf}$ ranging from close to zero (strong conformity) to unity (no conformity).}\label{fig:e_s_1D_conformity_efficiency}
\end{figure}

The effect of conformity, parameterized by $\xi_{\rm conf}$, is shown in Figure \ref{fig:e_s_1D_conformity_efficiency} and is obtained by dividing the blue lines in Figure \ref{fig:e_s_1D_conformity} by the red lines. Although we are looking at a highly derived quantity, the details of which should therefore be treated with some caution, it is noticeable that there is a large variation in the strength of conformity, with values of $\xi_{\rm conf}$ close to zero (strong conformity) in some regions of parameter space and close to unity (no conformity) in others.  

The strength of the conformity effect appears to weaken as we increase the halo mass from $\log m_{\rm h}  \sim 12.5$ to $\log m_{\rm h} \sim 13.5$ and beyond.   How could this arise?  One possibility (but assuredly not the only one) would be that in lower mass halos around $\log m_{\rm h}  \sim 12.5$, quenching is caused by a ``hidden common parameter'' associated with, say, a cut-off in gas supply that affects both centrals and satellites, so that conformity would be strong.  If, however the presence of star formation in the very few star-forming centrals in higher mass halos $\log m_{\rm h}  \sim 13.5$ was unrelated to a common cut-off of fuel, perhaps being rejuvenated during a merger of the central galaxy, then we would expect conformity to be much weaker or absent.  

This example emphasizes a fundamental difficulty with understanding conformity effects.  It is already difficult enough to interpret trends in $\epsilon_{\rm sat}$ with our different parameters because of built-in correlations between those parameters.  Interpreting trends in the strength of conformity $\xi_{\rm conf}$ with those same parameters is even harder because we are dealing with the effects of ``hidden variables'' whose identity is by definition unknown.  For this reason we do not speculate further about the physical origin of some of the other trends, e.g., the increase in the strength of the conformity with increasing stellar mass of the satellite.

In the context of the discussion in Section \ref{sec:meaning_conformity}, it should be noted that the ``one-dimensional'' conformity strength plots in each panel of Figure \ref{fig:e_s_1D_conformity_efficiency} give the strength of the conformity that is computed when matching only the single parameter plotted on the horizontal axis in each panel.  This is because they are computed using the average values of $\epsilon_{\rm sat}$ from the corresponding panels in Figure \ref{fig:e_s_1D_conformity}.  However, the close similarity noted in Section \ref{sec:current_conformity} between the overall conformity that is obtained from the average $\epsilon_{\rm sat}$ values for all satellites and that obtained from the highly restrictive subsample matched in all five parameters suggests that this may not be an important distinction.

\section{Discussion} \label{sec:discussion}

In the Sections \ref{sec:centrals_satellites}-\ref{sec:galaxy_conformity}, we have developed our understanding of the environmental effects on galaxies in groups within the context of the ``quenching efficiency'' formalism of galaxies, in which the quenched fraction of a set of galaxies is compared with a control sample; in this case the general set of centrals of the same stellar mass, which are overwhelmingly dominated by singleton centrals.  At least in the case of satellites, this control sample is presumably representative of what the galaxies would have been before they entered the group environment.

The main observational results presented in this paper can be summarized as follows.

First, the centrals of groups experience similar environmental effects as their satellites relative to the control sample of general centrals, i.e., the well-known distinction between centrals and satellites in their quenched fractions disappears when we consider the centrals of the groups containing the satellites. This result holds at least for the range of stellar mass that is sampled by the centrals within the groups, i.e.,  $m_\ast \gtrsim 10^{10.3}\: M_\odot$. While it is true that environmental effects within the galaxy population are dominated by satellites \citep[e.g.,][]{vandenbosch2008, peng2012, knobel2013, kovac2014} this is because, within these groups, satellites far outweigh the single central per group.     The demonstration that centrals also feel the same environmental quenching effects as satellites suggests that we can replace the idea of satellite quenching with ``group quenching''.    

Second, we have examined the dependence of the quenching of satellites on six parameters by looking first at the dependence of $\epsilon_{\rm sat}$ on the halo mass $m_{\rm h}$, the overdensity $\delta$, the group-centric distance $R$, the stellar masses of the satellite, $m_{\rm sat}$, and its central, $m_{\rm cen}$, and the sSFR of the central.  The first three of these all effect $\epsilon_{\rm sat}$ quite strongly, but these parameters are inter-related in the data. We therefore also looked at the variation of $\epsilon_{\rm sat}$ in two-dimensional plots, using pairs of these parameters.  Generally, we find similar results to what has been seen before, i.e., $\delta$, $R$, and $m_{\rm h}$ are all playing a role in the quenching of satellites.  It is hard to clearly disentangle the effects of one from the other, and we can safely conclude only that these three have a much stronger effect on $\epsilon_{\rm sat}$ than either the stellar mass of the satellite or that of the central.

Third, we have explored in more detail than hitherto the meaning and occurrence of ``galactic conformity'', i.e., the effect by which the quenching of satellites and centrals are correlated.  This is a strong effect in the sense that satellites of star forming centrals are much less likely to be quenched than the very much larger population of satellites of quenched centrals.  We have argued that the presence of conformity in a given sample and analysis is the reflection of the action of ``hidden common variables'' that (1) must affect the quenching of both satellites and centrals, (2) must somehow be ``shared'' by the centrals and satellites in a given group, and (3) must not have been an ``observable'' in the sense of having been controlled in the analysis, e.g., through matching (or binning) of that variable in the samples.

We have only been able to probe these environmental effects within the halo mass range in the \cite{yang2012} SDSS group catalog that we have been using, i.e., $\log m_{\rm h} \gtrsim 12.5$.  The mean $\epsilon_{\rm sat}$ in all groups is falling as we approach this limit (from above), and has a value $\epsilon_{\rm sat} \sim 0.2$ at $\log m_{\rm h} = 12.5$, only a third of that found at $\log m_{\rm h} = 14.5$.   The effects of conformity are however strong at the lower end of this mass range, and the satellites of star-forming centrals at $\log m_{\rm h} = 12.5$ already show no sign of environmental quenching.  It should be noted that the rise of $\epsilon_{\rm sat}$ with halo mass is not driven by the changing fraction of star-forming to quenched centrals as the halo mass increases.  The increase of $\epsilon_{\rm sat}$ with halo mass is seen in the satellites of both quenched and star-forming centrals and the decreasing importance of the star-forming centrals at higher halo masses has a negligible effect on the overall population because there are so few of them.

Over the whole halo mass range, there is evidence for a strong radial (or local overdensity) dependence of $\epsilon_{\rm sat}$, but not much evidence that the effects of conformity vary strongly with group-centric distance and overdensity.   Satellite galaxies appear to experience as strong a conformity effect out to $\sim \! 2R$, where the group-centric distance $R$ is defined in Equation (\ref{eq:R}).  As far as we can tell, the effect of conformity extends across much of the halo, or at least affects most of the group members --- we note that on average the virial radius typically corresponds to $2 \lesssim R \lesssim 3$.

Our whole analysis has been based on the quenching efficiency formalism via $\epsilon_{\rm sat}$, which takes out the effects of mass quenching \citep{peng2010, peng2012} of the galaxies by comparing the quenched fraction to that of the set of centrals (dominated by singleton centrals) at a given stellar mass.  This approach not only allows us to see the environmental effects more clearly, but it is additionally physically motivated by the fact that these satellites were presumably once centrals of their own halos, and indeed in many cases will remain as ``centrals'' of their own subhalos.  

The observational evidence presented in this paper that centrals in the groups that contain the satellites also experience the same environmental effects as satellites opens the question of whether ``mass quenching'' and ``satellite quenching'' (or now more generally ``group quenching'') are actually the same thing, or at least closely related.

At first, they might be thought to be quite different.  A robust result \citep[][this work]{vandenbosch2008,peng2010,peng2012} has been that the value of $\epsilon_{\rm sat}$, i.e., the strength of satellite quenching, is strikingly independent of the stellar mass of the satellite.  In contrast, mass quenching is, by definition, strongly dependent on the stellar mass \citep{peng2012}.   Indeed, our whole analysis has been based on the fact that the effects of stellar mass and of environment are multiplicative, i.e., that the overall quenched fraction $f_{\rm q}$ is separable in the two dimensions of mass and environment \citep{peng2010}.

However, it should be appreciated that many of the quantities that might be associated with the quenching of satellites will, to first order, be independent of the stellar mass of the satellite.  Put another way, at a given satellite stellar mass, the distributions of these variables across the satellite population will not depend strongly on the choice of the stellar mass.  Examples of such variables would include the halo mass, the group-centric distance, overdensity (assuming there is no mass segregation in the group), and the time of infall. Indeed the first three of these are precisely the variables that we and others have shown most strongly affect the strength of satellite quenching.   If satellite quenching is driven by one or more of these, then we would not expect $\epsilon_{\rm sat}$ to vary with stellar mass, provided (and this is an important and interesting caveat) the ability of a satellite to resist quenching and continue forming stars is not itself a function of its stellar mass.  

\begin{figure}
	\centering
	\includegraphics[width=0.4\textwidth]{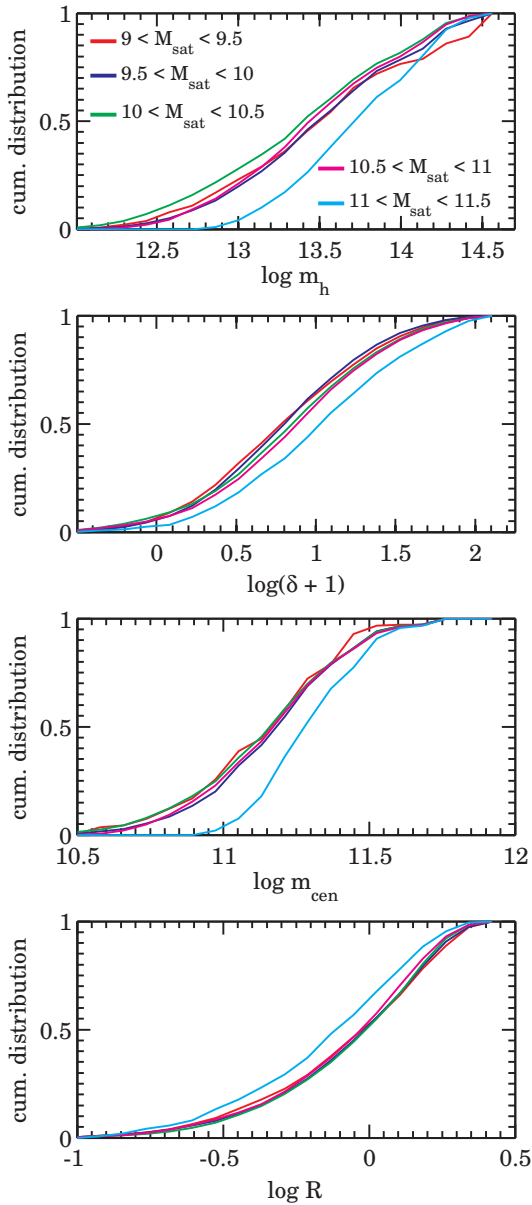}
	\caption{Cumulative distributions of the environmental parameters for our satellite sample in different stellar mass bins. Each color corresponds to a stellar mass bin as indicated in the top panel. It is obvious that the distributions are fairly similar for different mass bins except for the highest mass bin (i.e., $\log m_{\rm sat} > 11$). To compute these distributions the satellites are weighed (see Section \ref{sec:basic_section}).}\label{fig:satellite_distribution}
\end{figure}

That the distributions of these parameters are indeed more or less independent of stellar mass within the current sample is shown in Figure \ref{fig:satellite_distribution} which plots the cumulative distributions of $m_{\rm h}$, $m_{\rm cen}$, $\delta$, and $R$ for the five bins of satellite stellar mass used in Figure \ref{fig:e_s_1D}.  Only the most massive satellites with $\log m_{\rm sat} > 11$ have significantly different distributions in these parameters, for the obvious reason that these very massive satellites will not exist in the lowest mass halos.  This difference in the distribution of $m_{\rm h}$ for the most massive satellites then folds through to differences in $\delta$ because of the correlation between these parameters.  The difference in the $R$ distribution could reflect some mass segregation within the groups.

Suggestive evidence that mass and environment quenching may be very closely linked comes from the convergence of the halo mass at which these effects occur. Satellite quenching starts to become important around $11.5 \lesssim \log m_{\rm h} \lesssim 12.5$, when the curves of $\epsilon_{\rm sat}(m_{\rm h})$ (linearly extrapolated) intercept the $\epsilon_{\rm sat} = 0$ axis in Figure \ref{fig:e_s_1D_conformity}. Thus, the onset of environmentally driven satellite quenching occurs at more or less the halo mass that is associated
\citep[from abundance matching, e.g.,][]{moster2013,behroozi2013} with the Schechter $M^\ast$ in stellar mass (i.e., $\log m_{\ast} \sim 10.8$) that characterizes mass quenching \citep{peng2010,peng2012}.

Observational evidence in favor of linking the two effects comes, for example, from the recent structural analysis of \cite{carollo2014} who pointed out that the structural mix of quenched satellites does not depend on the group-centric distance, even though the fraction of these quenched satellites that can be inferred to have been environment-quenched (as opposed to mass-quenched) changes substantially with distance, from essentially zero on the group outskirts to about one-third at the center of the group. This was taken to indicate that either the structural and morphological signatures of the two putative quenching channels are identical, or that neither is associated with a structural or morphological signature.  In other words, the overall morphology-density relation is primarily a reflection of $\epsilon_{\rm sat}$ and not of variations within the quenched population due to different quenching channels (see \citealt{carollo2014} and discussion therein).

Linking the quenching of centrals and satellites in this way, and the evident strong effect of conformity (especially around the halo masses at which quenching appears) that we have highlighted in this paper, starts to strongly constrain the physical origin of quenching.   Recall that ``conformity'' is a reflection of the action of a hidden common variable that affects the quenching of both satellites and centrals, has a value that is ``shared'' between a central and its satellites. Furthermore, the quenching effect comes from a component of the hidden variable that is not correlated with any of the controlled parameters (five in the current study, including halo mass).  

Taken at face value, this would seem to argue against processes that are internal to individual galaxies, including internal structural changes and AGN activity, unless that information could be transmitted across the halo via a ``hidden common variable''.  Even then, we reiterate that the trend in $\epsilon_{\rm sat}$ with halo mass in Figure \ref{fig:e_s_1D_conformity} is not simply due to the effect of the changing fraction of quenched and star-forming centrals. In other words, some kind of ``quenching at a distance'' by an internal process operating in the central (or satellites) is not enough, because both $\epsilon_{\rm sat,q}$ and $\epsilon_{\rm sat,SF}$ increase with $m_{\rm h}$ as well as with $\delta$ and $R$.  

If the response of centrals and satellites to the environment is after all the same, then this would also suggest that any processes that would be unique to satellites, or to centrals, are unlikely to be dominating the environmental response of either.  A significant caveat to this is that we have only been able to establish this similarity over a very small range of stellar mass.

\section{Conclusions}\label{sec:conclusion}

In this paper, we have re-examined the quenched fractions of galaxies in the SDSS DR7 spectroscopic catalog as a function of their group environment.   We consider both the satellites and the centrals in the groups.  The analysis was performed in the framework of the ``quenching efficiency'' formalism that removes the large effects of stellar mass on the quenched fractions through comparison with a control sample of galaxies at the same stellar mass. This control sample is the set of central galaxies that are mostly singletons and which can be considered to be representative of the likely progenitors of the satellites in the groups.  

The principal findings in this paper are as follows.
\begin{enumerate}
\item Whereas, at a given stellar mass, satellites are systematically more quenched than the control sample of general centrals (as has been seen before), we show that the centrals of these same groups are also more quenched than the control sample.  In fact, as far as we can see, the centrals in these groups are experiencing essentially the same environmental quenching effects as the satellites when we compare centrals and satellites with the same environmental parameters, e.g., overdensity, group-centric distance, and halo mass. This suggests that the concept of ``satellite quenching'' that was introduced earlier \citep[e.g.,][]{vandenbosch2008,peng2012}, and which is describable by a satellite quenching efficiency $\epsilon_{\rm sat}(p_1,p_2,\ldots)$, should be generalized to ``group quenching'' that affects centrals and satellites in the same way (at least within the mass range that we can probe, i.e., $m_\ast \gtrsim 10^{10.3}\: M_\odot$).

\item The dependence of $\epsilon_{\rm sat}$ on different parameters for the satellites shows a wide range of behaviors. As has been seen before, $\epsilon_{\rm sat}$ is independent of the stellar mass of the satellite (we show it is also independent of the stellar mass of the central) but depends strongly on the inter-linked parameters of overdensity $\delta$, the group-centric distance $R$ with also a dependence on halo mass $m_{\rm h}$.  It is hard to isolate the effects of these, even on two-dimensional plots, and no simple dependences emerge across the whole of parameter space.  Some of the difficulty may stem from ambiguities in the definition of these parameters and/or uncertainties in their measurement.  

\item A very strong dependence is also seen on the sSFR of the central galaxy, a phenomenon called ``galactic conformity'' \citep{weinmann2006}.  A larger fraction of the satellites of quenched centrals are quenched than those of star-forming centrals.  This effect persists even when we carefully match the satellites in stellar mass $m_{\rm sat}$ and all four of the environmental parameters used in this study, $m_{\rm h}$, $m_{\rm cen}$, $\delta$, and $R$.  The persistence of conformity, even after such matching, is that the quenching of both the centrals and satellites must be strongly affected by one or more additional ``hidden common variables'' that have not been matched in the analysis and which contribute to the ``probabilistic'' aspect of quenching.  These hidden variables must also be in some way ``shared'' by the satellites and central of a given group, i.e., they must either be equal (as would be the case of a single parameter describing the group) or at least correlated (e.g., as in local overdensity).  The strength of conformity can be described by the ratio $\xi_{\rm conf}$ of the $\epsilon_{\rm sat}$ for those satellites with star-forming and quenched centrals.

\item This conformity strength $\xi_{\rm conf}$ has an average value of about 0.4, meaning that satellites of quenched centrals respond to the environment $\sim \! 2.5$ times more strongly than do those of star forming centrals.  This has the same value whether we simply average over all satellites or look at the five-parameter-matched sample.   However, the value of $\xi_{\rm conf}$ varies with most of the parameters and has values over the whole interval between zero (very strong conformity) and almost unity (no conformity) across the parameter space.  Conformity is strong at low halo masses (i.e., $m_{\rm h} \sim 10^{12.5}\: M_\odot$) but much weaker at high halo masses.  However, a variation in the strength of conformity can have several causes and so we do not attempt to interpret these trends in detail. It is a curious coincidence that this mean value of $\xi_{\rm conf}$ is essentially the same as the mean value of $\epsilon_{\rm sat}$.

\item A very interesting question raised by these results is whether our new environmentally driven group quenching process could be essentially the same process as that which drives the mass quenching \citep{peng2010,peng2012} that we have up to this point separated out from the analysis through the use of the $\epsilon_{\rm sat}$ formalism. It was suggested in \cite{carollo2014} that these could be different manifestations of the same underlying process, based on the observation that there seems to be no detectable difference in the structural properties of galaxies that were inferred to have undergone mass quenching and environment quenching and that the apparent mass independence of satellite quenching could then arise if the distribution of host halo masses for satellites was largely independent of their stellar mass, as indicated by the independence of the satellite mass function on group halo mass (e.g., Peng et al.\ 2012).  We showed here that the distribution of all of the environmental parameters that are most important for satellite quenching, i.e., $m_{\rm h}$, $\delta$, and $R$ are more or less independent of the stellar mass of a satellite (except for the most massive satellites which are not found in low mass halos). This would require that the response of a satellite to an environmental quenching ``pressure'' must be interestingly independent of its mass.

A strong argument in favor of linking environment quenching and mass quenching is that the halo mass scale at which environment quenching is inferred to start, which we obtain by a small extrapolation of $\epsilon_{\rm sat}(m_{\rm h})$, is very similar to the halo mass that corresponds to the stellar mass that is associated with mass quenching.

\item The fact that the strong galactic conformity effect linking the quenching of centrals and satellites extends (without a strong variation in strength) out to group-centric distances of 2-3 $R$, i.e., comparable to the virial radius, suggests that the causes and/or effects of quenching cannot be local within galaxies and must rather extend across the whole extent of halos.  This argues against internal galaxy processes driving quenching, unless the subsequent effects can be transmitted over large distances.  Even then, it is clear that the quenching of the central alone is not sufficient (even with conformity) to explain the increase of satellite quenching with e.g., halo mass, since this is observed separately for the satellites of both quenched and star-forming centrals.

Another interesting constraint is that, if the response of centrals and satellites to the environment is the same, as we have suggested here at least in the range of parameter space sampled by both, then it argues against the importance of satellite-specific physical processes in the quenching of galaxies.  The validity of this conclusion at low stellar masses is, however, conjectural and is based only on the mass independence of the environmental quenching of satellites.

\end{enumerate}

This research was supported by the Swiss National Science Foundation. The idea that environmental quenching and mass quenching could be closely related was stimulated by a conversation with John Kormendy, which we gratefully acknowledge. We also thank Marcella Carollo for helpful discussions and the anonymous referee for their careful reading of the paper.

\bibliographystyle{apj3}
\bibliography{apj-jour,bibliography}

\end{document}